\documentclass[twocolumn,pre,aps,showpacs,showkeys,amsmath]{revtex4}
\usepackage{graphicx}

\begin{document}

\title{Effect of Inhibitory Spike-Timing-Dependent Plasticity on Fast Sparsely Synchronized Rhythms in A Small-World Neuronal Network}
\author{Sang-Yoon Kim}
\email{sykim@icn.re.kr}
\author{Woochang Lim}
\email{wclim@icn.re.kr}
\affiliation{Institute for Computational Neuroscience and Department of Science Education, Daegu National University of Education, Daegu 42411, Korea}

\begin{abstract}
We consider the Watts-Strogatz small-world network (SWN) consisting of inhibitory fast spiking Izhikevich interneurons. This inhibitory neuronal population has
adaptive dynamic synaptic strengths governed by the inhibitory spike-timing-dependent plasticity (iSTDP). In previous works without iSTDP, fast sparsely synchronized rhythms,
associated with diverse cognitive functions, were found to appear in a range of large noise intensities for fixed strong synaptic inhibition strengths. Here, we investigate the effect of iSTDP on fast sparse synchronization
(FSS) by varying the noise intensity $D$. We employ an asymmetric anti-Hebbian time window for the iSTDP update rule [which is in contrast to the Hebbian time window for the excitatory STDP (eSTDP)].
Depending on values of $D$, population-averaged values of saturated synaptic inhibition strengths are potentiated [long-term potentiation (LTP)] or depressed [long-term depression (LTD)] in comparison with the
initial mean value, and dispersions from the mean values of LTP/LTD are much increased when compared with the initial dispersion, independently of $D$.
In most cases of LTD where the effect of mean LTD is dominant in comparison with the effect of dispersion, good synchronization (with higher spiking measure) is found to get better via LTD, while bad
synchronization (with lower spiking measure) is found to get worse via LTP. This kind of Matthew effect in inhibitory synaptic plasticity is in contrast to that in excitatory synaptic plasticity where good (bad) synchronization gets better (worse) via LTP (LTD). Emergences of LTD and LTP of synaptic inhibition strengths are intensively investigated via a microscopic method based on the distributions of time delays between the pre- and the post-synaptic spike times. Furthermore, we also investigate the effects of network architecture on FSS by changing the rewiring probability $p$ of the SWN in the presence of iSTDP.
\end{abstract}


\maketitle

\section{Introduction}
\label{sec:INT}
In recent years, brain rhythms have attracted  much attention \cite{Buz1,TW,Rhythm5,Rhythm9,Rhythm4,Rhythm10,Rhythm8,Rhythm11,Rhythm3,Rhythm1,Rhythm12,Rhythm2,Rhythm6,Rhythm13,Rhythm7}. Particularly, we are interested in fast sparsely synchronized rhythms, associated with diverse cognitive functions (e.g., sensory perception, feature integration, selective attention, and memory formation) \cite{W_Review}. At the population level, fast sparsely synchronous oscillations [e.g., gamma rhythm (30-100 Hz) during awake behaving states and rapid eye movement sleep and sharp-wave ripple (100-200 Hz) during quiet sleep and awake immobility] have been observed in local field potential recordings, while at the cellular level individual neuronal recordings have been found to exhibit irregular and intermittent spike discharges like Geiger counters \cite{SS1,SS3,SS2,SS4,SS5,SS6,SS7}. Thus, single-cell firing activity differs distinctly from the population oscillatory behavior. We note that these fast sparsely synchronized rhythms are in contrast to fully synchronized rhythms where individual neurons fire regularly at the population frequency like clocks. Under the balance between strong external noise and strong recurrent inhibition, fast sparse synchronization (FSS) was found to appear in both random networks \cite{Sparse1,Sparse2,Sparse3,Sparse4} and globally-coupled networks \cite{Sparse5,Sparse6}. In brain networks, architecture of synaptic connections has been found to have complex topology (e.g., small-worldness and scale-freeness) which is neither regular nor completely random \cite{CN6,CN1,CN2,CN7,CN3,CN4,CN5,Sporns}. In our recent works \cite{FSS-SWN,FSS-SFN}, as complex networks we employed the small-world network and the scale-free network, and studied the effects of network architecture on emergence of FSS.

In the previous works on FSS, synaptic coupling strengths were static. However, in real brains synaptic strengths may change to adapt to the environment
(i.e., they can be potentiated \cite{LTP2,LTP1,LTP3} or depressed \cite{LTD2,LTD1,LTD3,LTD4}). These adjustments of synapses are called the synaptic plasticity which provides the basis for
learning, memory, and development \cite{Abbott1}. Regarding the synaptic plasticity, we consider a spike-timing-dependent plasticity (STDP) \cite{STDP1,STDP2,STDP3,STDP4,STDP5,STDP6,STDP7,STDP8}.
For the STDP, the synaptic strengths vary via an update rule depending on the relative time difference between the pre- and the post-synaptic spike times.
Many models for STDP have been employed to explain results on synaptic modifications occurring in diverse neuroscience topics for health and disease (e.g., temporal sequence learning \cite{TSLearning}, temporal pattern recognition \cite{EtoE6}, coincidence detection \cite{EtoE0}, navigation \cite{Navi}, direction selectivity \cite{DirSel}, memory consolidation \cite{Memory}, competitive/selective development \cite{Devel}, and deep brain stimulation \cite{Lou}). Recently, the effects of STDP on population synchronization in ensembles of coupled neurons were also studied in various aspects \cite{Tass1,Tass2,Brazil1,Brazil2,SBS,SSS}.

A neural circuit in the brain cortex is composed of a few types of excitatory principal cells and diverse types of inhibitory interneurons.
These interneurons make up about 20 percent of all cortical neurons, and exhibit diversity in their morphologies and functions \cite{Buz2}.
By providing a synchronous oscillatory output to the principal cells, interneuronal networks play the role of backbones of many cortical rhythms \cite{GR,WB,Wang,Rhythm5}.
Synaptic plasticity of excitatory and inhibitory connections is of great interest because it controls the efficacy of potential computational functions of excitation and inhibition.
Studies of synaptic plasticity have been mainly focused on synaptic connections between excitatory pyramidal cells, because excitatory-to-excitatory (E to E) connections are most prevalent in the cortex
and they form a relatively homogeneous population \cite{EtoE0,EtoE1,EtoE3,EtoE4,EtoE2,EtoE6,EtoE5,EtoE7,EtoE8}. An asymmetric Hebbian time window was employed for the excitatory STDP (eSTDP) update rule \cite{STDP1,STDP2,STDP3,STDP4,STDP5,STDP6,STDP7,STDP8}. When a pre-synaptic spike precedes a post-synaptic spike, long-term potentiation (LTP) occurs; otherwise, long-term depression (LTD) appears.
On the other hand, plasticity of inhibitory connections has attracted less attention mainly because of experimental obstacles and diversity of interneurons \cite{iSTDP4,iSTDP3,iSTDP2,iSTDP1,iSTDP5}.
With the advent of fluorescent labeling and optical manipulation of neurons according to their genetic types \cite{iExpM1,iExpM2}, inhibitory plasticity has also begun to be focused.
Particularly, studies on iSTDP of inhibitory-to-excitatory (I to E) connections  have been much made. Thus, inhibitory STDP (iSTDP) has been found to be diverse and cell-specific \cite{iSTDP12,iSTDP4,iSTDP11,iSTDP10,iSTDP8,iSTDP3,iSTDP7,iSTDP6,iSTDP2,iSTDP1,iSTDP5,iSTDP9}.

In this paper, we consider an inhibitory Watts-Strogatz small-world network (SWN) of fast spiking (FS) interneurons \cite{SWN1,SWN2,SWN3}, and investigate the effect of iSTDP [of inhibitory-to-inhibitory (I to I)
connections] on FSS by varying the noise intensity $D$. We employ an asymmetric anti-Hebbian time window for the iSTDP update rule, in contrast to the Hebbian time window for the eSTDP \cite{Tass1,Lou}.
Then, strengths of synaptic inhibition $\{ J_{ij} \}$ change with time, and eventually, they become saturated after a sufficiently long time.
Depending on $D$, mean values of saturated synaptic inhibition strengths $\{ J_{ij}^* \}$ are potentiated [long-term potentiation (LTP)] or depressed [long-term depression (LTD)], when compared with the initial mean value of synaptic inhibition strengths. On the other hand, dispersions from the mean values of LTP/LTD are much increased in comparison with the initial dispersion, irrespectively of $D$.

For the case of iSTDP, both the mean value and the dispersion (for the distribution of synaptic inhibition strengths) affect population synchronization.
The LTD (LTP) tends to increase (decrease) the degree of FSS due to decrease (increase) in the mean value of synaptic inhibition strengths, and the increased dispersions decrease the degree of FSS.
For most cases of LTD where the effect of mean LTD is dominant in comparison with the effect of dispersion, good synchronization (with higher spiking measure) gets better via LTD; in some other cases where dispersions are dominant, the degree of good synchronization may be decreased even in the case of LTD. On the other hand, in all cases bad synchronization (with lower spiking measure) gets worse via LTP. This kind of Matthew effect (valid in most cases of LTD) is in contrast to that in the case of eSTDP where good (bad) synchronization gets better (worse) via LTP (LTD) \cite{SSS,SBS}; the role of LTD (LTP) in the case of iSTDP is similar to that of LTP (LTD) for the case of eSTDP. Emergences of LTD and LTP of synaptic inhibition strengths are also investigated through a microscopic method based on the distributions of time delays between the
nearest spiking times of the pre- and the post-synaptic interneurons. Moreover, we study the effects of network architecture on FSS by varying the rewiring probability $p$ of the SWN in the presence of iSTDP.

This paper is organized as follows. In Sec.~\ref{sec:SWN}, we describe an inhibitory Watts-Strogatz SWN of FS interneurons with inhibitory synaptic plasticity.
Then, in Sec.~\ref{sec:FSS} the effects of iSTDP on FSS are investigated. Finally, we give summary and discussion in Sec.~\ref{sec:SUM}.

\section{Watts-Strogatz SWN of FS Izhikevich Interneurons with Inhibitory Synaptic Plasticity}
\label{sec:SWN}
We consider an inhibitory directed Watts-Strogatz SWN, composed of $N$ FS interneurons equidistantly placed on a one-dimensional ring of radius $N/ 2 \pi$. The
Watts-Strogatz SWN interpolates between a regular lattice with high clustering (corresponding to the case of $p=0$) and a random graph with short average path length (corresponding to the case
of $p=1$) via random uniform rewiring with the probability $p$ \cite{SWN1,SWN2,SWN3}. For $p=0,$ we start with a directed regular ring lattice with $N$ nodes where each node is coupled to its first $M_{syn}$ neighbors ($M_{syn}/2$ on either side) via outward synapses, and rewire each outward connection uniformly at random over the whole ring with the probability $p$ (without self-connections and duplicate connections).
This Watts-Strogatz SWN model may be regarded as a cluster-friendly extension of the random network by reconciling the six degrees of separation (small-worldness) \cite{SDS1,SDS2} with the circle of friends (clustering).
These SWNs with predominantly local connections and rare long-range connections were employed in many recent works on various subjects of neurodynamics \cite{SW2,SW3,SW4,SW5,SW6,SW7,SW8,SW9,SW10,SW11,SW12,SW13}.

\begin{table}
\caption{Parameter values used in our computations; units of the capacitance, the potential, the current, and the time are pF, mV, pA, and msec, respectively.}
\label{tab:Parm}
\begin{ruledtabular}
\begin{tabular}{llllll}
(1) & \multicolumn{5}{l}{Single Izhikevich Fast Spiking Interneurons \cite{Izhi3}} \\
&  $C=20$ & $v_{r}=-55$ & $v_{t}=-40$ & $v_{p}=25$ & $v_{b}=-55$ \\
&  $k=1$ & $a=0.2$ & $b=0.025$ & $c=-45$ & $d=0$ \\
\hline
(2) & \multicolumn{5}{l}{Random External Excitatory Input to Each Izhikevich}\\
& \multicolumn{5}{l}{Fast Spiking Interneurons} \\
& \multicolumn{2}{l}{$I_{i} \in [680, 720]$} & \multicolumn{3}{l}{$D$: Varying} \\
\hline
(3) & \multicolumn{5}{l}{Inhibitory Synapse Mediated by The GABA$_{\rm A}$}\\
& \multicolumn{5}{l}{Neurotransmitter \cite{Sparse3}} \\
& $\tau_l=1$ & $\tau_r=0.5$ & $\tau_d=5$ & \multicolumn{2}{l}{$V_{syn}=-80$} \\
\hline
(4) & \multicolumn{5}{l}{Synaptic Connections between Interneurons in The}\\
& \multicolumn{5}{l}{Watts-Strogatz SWN} \\
& $M_{syn}=50$ & \multicolumn{4}{l}{$p$: Varying} \\
& $J_{0}=700$ & $\sigma_0=5$  & \multicolumn{3}{l}{$J_{ij} \in [0.0001, 2000]$} \\
\hline
(5) & \multicolumn{5}{l}{Anti-Hebbian iSTDP Rule} \\
& $\delta = 0.05$ & $A_{+} = 1.0$ & $A_{-} = 1.1$ & $\tau_{+} = 11.5$ & $\tau_{-} = 12$ \\
\end{tabular}
\end{ruledtabular}
\end{table}

As elements in our SWN, we choose the Izhikevich inhibitory FS interneuron model which is not only biologically plausible, but also computationally efficient \cite{Izhi1,Izhi2,Izhi3,Izhi4}.
Unlike Hodgkin-Huxley-type conductance-based models, the Izhikevich model matches neuronal dynamics by tuning the parameters instead of matching neuronal electrophysiology. The parameters $k$ and $b$ are associated with the neuron's rheobase and input resistance, and $a,$ $c$, and $d$ are the recovery time constant, the after-spike reset value of $v$, and the after-spike jump value of $u$, respectively. Tuning these parameters, the Izhikevich neuron model may produce 20 of the most prominent neuro-computational features of biological neurons \cite{Izhi1,Izhi2,Izhi3,Izhi4}.
Particularly, the Izhikevich model is used to reproduce the six most fundamental classes of firing patterns observed in the mammalian neocortex; (a) excitatory regular spiking pyramidal neurons, (b) inhibitory FS interneurons, (c) intrinsic bursting neurons, (d) chattering neurons, (e) low-threshold spiking neurons and (f) late spiking neurons \cite{Izhi3}.
Here, we use the parameter values for the FS interneurons (which do not fire postinhibitory rebound spikes) in the layer 5 rat visual cortex, which are listed in the 1st item of Table \ref{tab:Parm}
(see the caption of Fig. 8.27 in \cite{Izhi3}).

The following equations (\ref{eq:PD1})-(\ref{eq:PD6}) govern the population dynamics in the SWN:
\begin{eqnarray}
C\frac{dv_i}{dt} &=& k (v_i - v_r) (v_i - v_t) - u_i +I_{i} \nonumber \\ && +D \xi_{i} -I_{syn,i}, \label{eq:PD1} \\
\frac{du_i}{dt} &=& a \{ U(v_i) - u_i \},  \;\;\; i=1, \cdots, N, \label{eq:PD2}
\end{eqnarray}
with the auxiliary after-spike resetting:
\begin{equation}
{\rm if~} v_i \geq v_p,~ {\rm then~} v_i \leftarrow c~ {\rm and~} u_i \leftarrow u_i + d, \label{eq:PD3}
\end{equation}
where
\begin{eqnarray}
U(v) &=& \left\{ \begin{array}{l} 0 {\rm ~for~} v<v_b \\ b(v - v_b)^3 {\rm ~for~} v \ge v_b \end{array} \right. , \label{eq:PD4} \\
I_{syn,i} &=& \frac{1}{d_i^{(in)}} \sum_{j=1 (j \ne i)}^N J_{ij} w_{ij} s_j(t) (v_i - V_{syn}), \label{eq:PD5}\\
s_j(t) &=& \sum_{f=1}^{F_j} E(t-t_f^{(j)}-\tau_l); \nonumber \\
E(t) &=& \frac{1}{\tau_d - \tau_r} (e^{-t/\tau_d} - e^{-t/\tau_r}) \Theta(t). \label{eq:PD6}
\end{eqnarray}
Here, the state of the $i$th neuron at a time $t$ is characterized by two state variables: the membrane potential $v_i$ and the recovery current $u_i$. In Eq.~(\ref{eq:PD1}), $C$ is the membrane capacitance, $v_r$ is the resting membrane potential, and $v_t$ is the instantaneous threshold potential. After the potential reaches its apex (i.e., spike cutoff value) $v_p$, the membrane potential and the recovery variable are reset according to Eq.~(\ref{eq:PD3}). The units of the capacitance $C$, the potential $v$, the current $u$ and the time $t$ are pF, mV, pA, and msec, respectively.
The parameter values used in our computations are listed in Table \ref{tab:Parm}. More details on the random external excitatory input to each Izhikevich FS interneuron, the synaptic currents and plasticity,
and the numerical method for integration of the governing equations are given in the following subsections.

\begin{figure}[b]
\includegraphics[width=0.9\columnwidth]{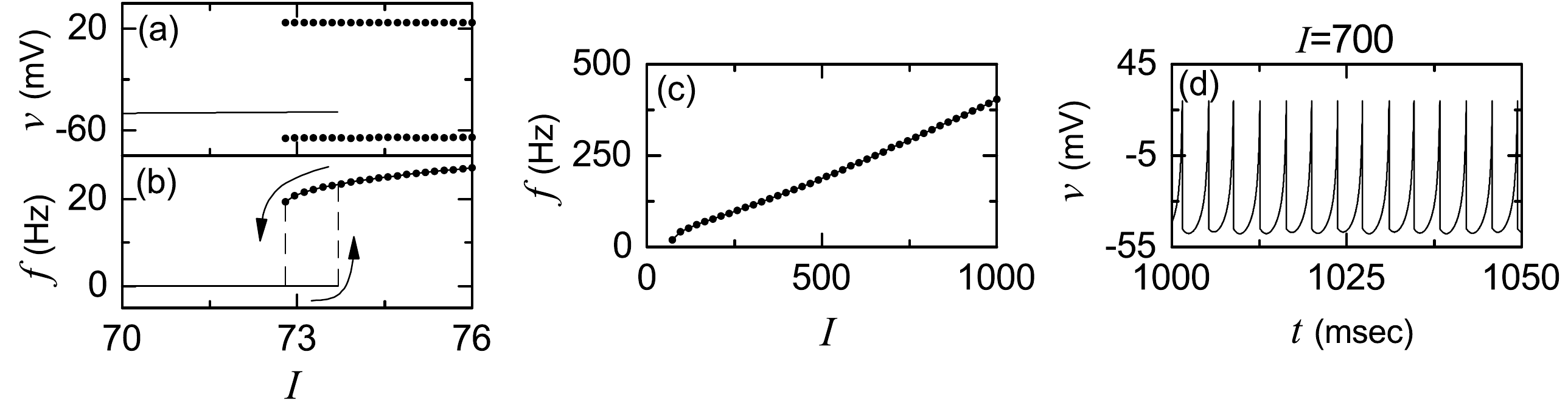}
\caption{Single FS Izhikevich interneuron for $D=0$.
(a) Bifurcation diagram. Solid line denotes a stable equilibrium point, and maximum and minimum values of $v$ for the spiking state are represented by solid circles.
Plot of the mean firing rate $f$ versus the time-averaged constant $I$ of the random external excitatory input (b) near the transition point and (c) in a large range of $I$. (d) Time series of the membrane potential $v$
for $I=700$.
}
\label{fig:Single}
\end{figure}

\subsection{Random External Excitatory Input to Each Izhikevich FS Interneuron}
\label{subsec:Sti}
Each interneuron in the network receives stochastic external excitatory input $I_{ext,i}$ from other brain regions, not included in the network (i.e., corresponding to background excitatory input)
\cite{GR, Sparse1,Sparse2,Sparse3,Sparse4}. Then, $I_{ext,i}$ may be modeled in terms of its time-averaged constant $I_i$ and an independent Gaussian white noise $\xi_i$ (i.e., corresponding to fluctuation of
$I_{ext,i}$ from its mean) [see the 3rd and the 4th terms in Eq.~(\ref{eq:PD1})] satisfying $\langle \xi_i(t) \rangle =0$ and $\langle \xi_i(t)~\xi_j(t') \rangle = \delta_{ij}~\delta(t-t')$, where $\langle\cdots\rangle$ denotes the ensemble average. The intensity of the noise $\xi_i$ is controlled by using the parameter $D$. In the absence of noise (i.e., $D=0$), the Izhikevich interneuron exhibits a jump from a resting state to a spiking state via subcritical Hopf bifurcation for $I_h=73.7$ by absorbing an unstable limit cycle born via a fold limit cycle bifurcation for $I_l=72.8$, as shown in Fig.~\ref{fig:Single}(a) \cite{FSS-SWN}. Hence, the Izhikevich FS interneuron shows type-II excitability because it begins to fire with a non-zero frequency [see Fig.~\ref{fig:Single}(b)]  \cite{Ex1,Ex2}. As $I$ is increased from $I_h$, the mean firing rate (MFR) $f$ increases monotonically, as shown in Fig.~\ref{fig:Single}(c). Throughout this paper, we consider a suprathreshold case such that the value of $I_i$ is chosen via uniform random sampling in the range of [680,720], as shown in the 2nd item of Table \ref{tab:Parm}; for the middle value of $I=700$, the membrane potential $v$ oscillates very fast with $f=271$ Hz [see Fig.~\ref{fig:Single}(d)].

\subsection{Synaptic Currents and Plasticity}
\label{subsec:Syn}
The last term in Eq.~(\ref{eq:PD1}) represents the synaptic couplings of Izhikevich FS interneurons. $I_{syn,i}$ of Eq.~(\ref{eq:PD5}) represents a synaptic current injected into the $i$th interneuron, and $V_{syn}$ is the synaptic reversal potential. The synaptic connectivity is given by the connection weight matrix $W$ (=$\{ w_{ij} \}$) where  $w_{ij}=1$ if the interneuron $j$ is presynaptic to the interneuron $i$; otherwise, $w_{ij}=0$.
Here, the synaptic connection is modeled in terms of the Watts-Strogatz SWN. Then, the in-degree of the $i$th neuron, $d_i^{(in)}$ (i.e., the number of synaptic inputs to the interneuron $i$) is given by $d_i^{(in)} = \sum_{j=1 (j \ne i)}^N w_{ij}$. For this case, the average number of synaptic inputs per neuron is given by $M_{syn} = \frac{1}{N} \sum_{i=1}^{N} d_{i}^{(in)}$. Throughout the paper, $M_{syn}=50$ (see the 4th item of Table \ref{tab:Parm}).
The fraction of open synaptic ion channels at time $t$ is denoted by $s(t)$. The time course of $s_j(t)$ of the $j$th neuron is given by a sum of delayed double-exponential functions $E(t-t_f^{(j)}-\tau_l)$ [see Eq.~(\ref{eq:PD6})], where $\tau_l$ is the synaptic delay, and $t_f^{(j)}$ and $F_j$ are the $f$th spike and the total number of spikes of the $j$th interneuron at time $t$, respectively. Here, $E(t)$ [which corresponds to contribution of a presynaptic spike occurring at time $0$ to $s_j(t)$ in the absence of synaptic delay] is controlled by the two synaptic time constants: synaptic rise time $\tau_r$ and decay time $\tau_d$, and $\Theta(t)$ is the Heaviside step function: $\Theta(t)=1$ for $t \geq 0$ and 0 for $t <0$. For the inhibitory GABAergic synapse (involving the $\rm{GABA_A}$ receptors), the values of $\tau_l$, $\tau_r$, $\tau_d$, and $V_{syn}$ are listed in the
3rd item of Table \ref{tab:Parm} \cite{Sparse3}.

The coupling strength of the synapse from the $j$th pre-synaptic interneuron to the $i$th post-synaptic interneuron is $J_{ij}$.
Here, we consider a multiplicative iSTDP (dependent on states) for the synaptic strengths $\{ J_{ij} \}$ \cite{Multi,Tass2}.
To avoid unbounded growth and elimination of synaptic connections, we set a range with the upper and the lower bounds: $J_{ij} \in [J_l, J_h]$,
where $J_l=0.0001$ and $J_h=2000$. Initial synaptic strengths are normally distributed with the mean $J_0(=700)$ and the standard deviation $\sigma_0(=5)$. With increasing time $t$, the synaptic strength for each synapse is updated with a nearest-spike pair-based STDP rule \cite{SS}:
\begin{equation}
J_{ij} \rightarrow J_{ij} + \delta (J^*-J_{ij})~|\Delta J_{ij}(\Delta t_{ij})|,
\label{eq:MSTDP}
\end{equation}
where $\delta$ $(=0.05)$ is the update rate, $J^*=$ $J_h~(J_l)$ for the LTP (LTD), and $\Delta J_{ij}(\Delta t_{ij})$ is the synaptic modification depending on the relative time difference $\Delta t_{ij}$ $(=t_i^{(post)} - t_j^{(pre)})$ between the nearest spike times of the post-synaptic interneuron $i$ and the pre-synaptic interneuron $j$.
We use an asymmetric anti-Hebbian time window for the synaptic modification $\Delta J_{ij}(\Delta t_{ij})$ \cite{Tass1,Lou}:
\begin{equation}
  \Delta J_{ij}(\Delta t_{ij}) = \left\{ \begin{array}{l} -A_{+}~  e^{-\Delta t_{ij} / \tau_{+}} ~{\rm for}~ \Delta t_{ij} > 0\\
  - A_{-}~ \frac{\Delta t_{ij}}{\tau_{-}} ~ e^{\Delta t_{ij} / \tau_{-}} ~{\rm for}~ \Delta t_{ij} \le 0\end{array} \right. ,
\label{eq:TW}
\end{equation}
where $A_+=1.0$, $A_-=1.1$, $\tau_+=11.5$ msec, and $\tau_-=12$ msec (these values are also given in the 5th item of Table \ref{tab:Parm}).
For the case of $\Delta t_{ij} > 0$, LTD occurs, while LTP takes place in the case of $\Delta t_{ij} < 0$, in contrast to the Hebbian time window for the eSTDP where
LTP (LTD) occurs for $\Delta t_{ij} > (<) 0$ \cite{SSS}.

\subsection{Numerical Method for Integration}
\label{subsec:NM}
Numerical integration of stochastic differential Eqs.~(\ref{eq:PD1})-(\ref{eq:PD6}) with an anti-Hebbian iSTDP rule of Eqs.~(\ref{eq:MSTDP}) and (\ref{eq:TW})
is done by employing the Heun method \cite{SDE} with the time step $\Delta t=0.01$ msec. For each
realization of the stochastic process, we choose  random initial points $[v_i(0),u_i(0)]$ for the $i$th $(i=1,\dots, N)$ FS interneuron with uniform probability in the range of
$v_i(0) \in (-50,-45)$ and $u_i(0) \in (10,15)$.

\begin{figure}
\includegraphics[width=\columnwidth]{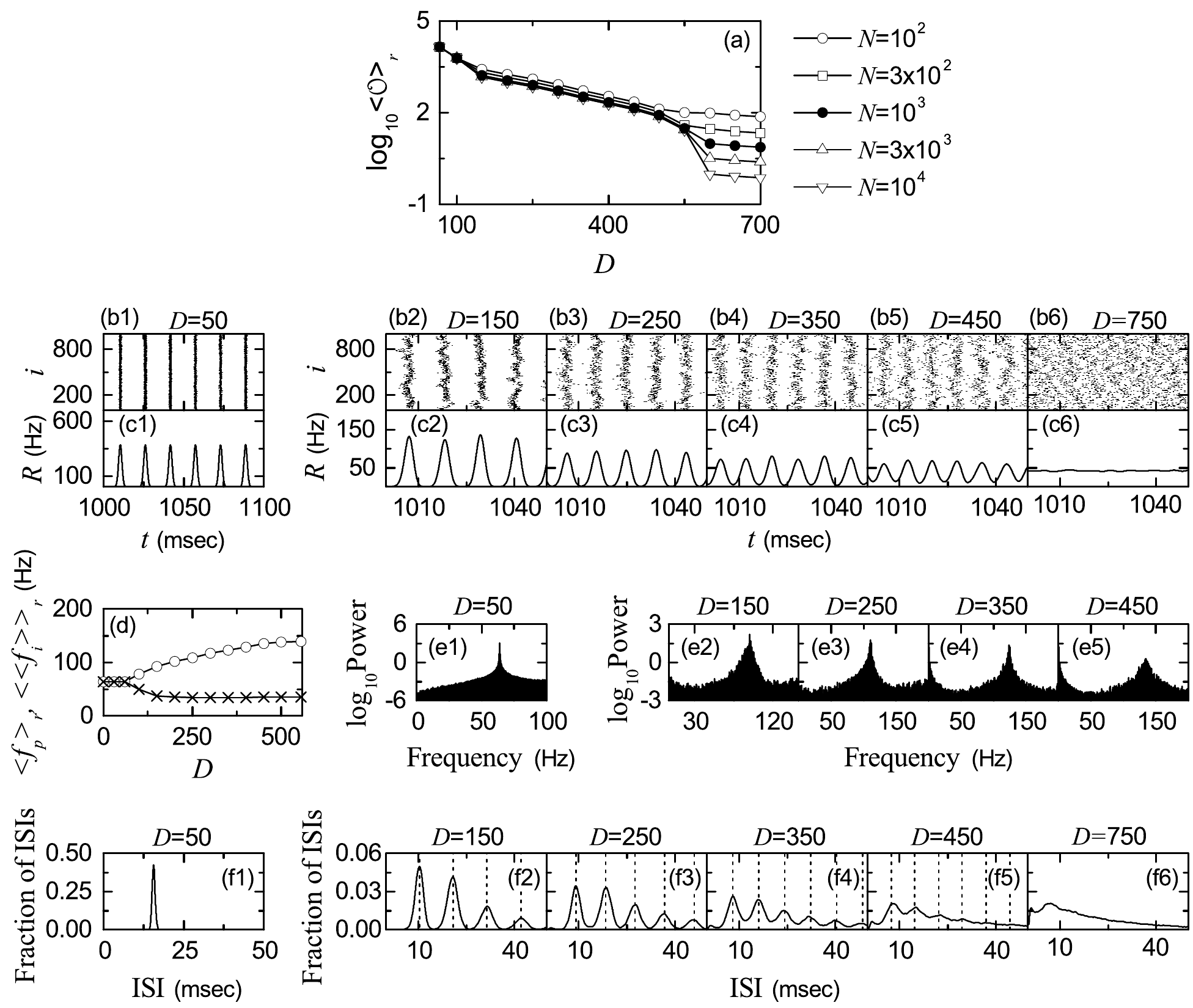}
\caption{Population states in the absence of iSTDP. (a) Plots of the thermodynamic order parameter $\langle {\cal{O}} \rangle_r$ versus $D$.
Raster plots of spikes in (b1)-(b6) and IPSR kernel estimates $R(t)$ in (c1)-(c6).
(d) Plots of the population frequency $\langle f_p \rangle_r$ (represented by open circles) and the population-averaged MFR of individual interneurons $\langle \langle f_i \rangle \rangle_r$ (denoted by
crosses) versus $D$. (e1)-(e5) Power spectra of $\Delta R(t) (= R(t)- \overline{R(t)})$ (the overbar represents the time average).
ISI histograms in (f1)-(f6); vertical dotted lines represent multiples of the global period $T_G$ of the IPSR $R(t)$.
}
\label{fig:NSTDP1}
\end{figure}

\section{Effect of Inhibitory STDP on Fast Sparsely Synchronized Rhythms}
\label{sec:FSS}
We consider the Watts-Strogatz SWN with high clustering and short path length when the rewiring probability $p$ is 0.25. This SWN is composed of $N$ inhibitory Izhikevich
FS interneurons. Throughout the paper, $N=10^3$ except for the cases in Fig.~\ref{fig:NSTDP1}(a) and in Figs.~\ref{fig:NSTDP3}(b1), \ref{fig:NSTDP3}(b2), and \ref{fig:NSTDP3}(c).

\subsection{FSS in The Absence of iSTDP}
\label{subsec:NSTDP}
First, we are concerned about FSS in the absence of iSTDP in the SWN with $p=0.25$.
The coupling strengths $\{ J_{ij} \}$ are static, and their values are chosen from the Gaussian distribution with the mean
$J_0$ (= 700) and the standard deviation $\sigma_0$ (=5). Population synchronization may be well visualized in the raster plot of neural spikes which is a collection of spike trains of individual interneurons. Such raster plots of spikes are fundamental data in experimental neuroscience. As a collective quantity showing population behaviors, we use an instantaneous population spike rate (IPSR) which may be obtained from the raster plots of spikes \cite{Sparse1,Sparse2,Sparse3,Sparse4,Sparse5,Sparse6,W_Review,RM}.
For the synchronous case, ``stripes" (composed of spikes and indicating population synchronization) are found to be formed in the raster plot, while in the desynchronized case spikes are completely scattered.
Hence, for a synchronous case, an oscillating IPSR $R(t)$ appears, while for a desynchronized case $R(t)$ is nearly stationary. To obtain a smooth IPSR, we employ the kernel density estimation (kernel smoother)
\cite{Kernel}. Each spike in the raster plot is convoluted (or blurred) with a kernel function $K_h(t)$ to obtain a smooth estimate of IPSR $R(t)$:
\begin{equation}
R(t) = \frac{1}{N} \sum_{i=1}^{N} \sum_{s=1}^{n_i} K_h (t-t_{s}^{(i)}),
\label{eq:IPSR}
\end{equation}
where $t_{s}^{(i)}$ is the $s$th spiking time of the $i$th interneuron, $n_i$ is the total number of spikes for the $i$th neuron, and we use a Gaussian
kernel function of band width $h$:
\begin{equation}
K_h (t) = \frac{1}{\sqrt{2\pi}h} e^{-t^2 / 2h^2}, ~~~~ -\infty < t < \infty.
\label{eq:Gaussian}
\end{equation}
Throughout the paper, the band width $h$ of $K_h(t)$ is 1 msec.

The mean square deviation of $R(t)$,
\begin{equation}
{\cal{O}} \equiv \overline{(R(t) - \overline{R(t)})^2},
 \label{eq:Order}
\end{equation}
plays the role of an order parameter $\cal{O}$ \cite{RM}; the overbar represents the time average. This order parameter may be regarded as a thermodynamic measure because it
concerns just the macroscopic IPSR kernel estimate $R(t)$ without any consideration between $R(t)$ and microscopic individual spikes. In the thermodynamic limit of
$N \rightarrow \infty$, the order parameter $\cal{O}$ approaches a non-zero (zero) limit value for the synchronized (desynchronized) state.
Figure \ref{fig:NSTDP1}(a) shows plots of $\log_{10} \langle {\cal O} \rangle_r$ versus $D$.
In each realization, we discard the first time steps of a stochastic trajectory as transients for $10^3$ msec, and then we numerically compute $\cal{O}$ by following the stochastic
trajectory for $3 \times 10^4$ msec. Throughout the paper, $\langle \cdots \rangle_r$ denotes an average over 20 realizations.
With increasing $N$ up to $10^4$, these numerical calculations for $\langle {\cal{O}} \rangle_r$ are done for various values of $D$.
For $D < D^*$($\simeq 558$), synchronized states exist because the order parameter $\langle {\cal{O}} \rangle_r$ tends to converge toward non-zero
limit values. On the other hand, for $D > D^*$, with increasing $N$ the order parameter $\langle {\cal{O}} \rangle_r$ tends to approach zero,
and hence a transition to desynchronization occurs due to a destructive role of noise spoiling the population synchronization.

Figures \ref{fig:NSTDP1}(b1)-\ref{fig:NSTDP1}(b6) show raster plots of spikes for various values of $D$, and their corresponding IPSR kernel estimates $R(t)$ are also shown in
Figs.~\ref{fig:NSTDP1}(c1)-\ref{fig:NSTDP1}(c6). For $D=50$, clear (straight) stripes appear successively in the raster plot of spikes, as shown in Fig.~\ref{fig:NSTDP1}(b1), and the corresponding IPSR $R(t)$
exhibits a regular oscillation [see Fig.~\ref{fig:NSTDP1}(c1)].
However, as $D$ is increased, the raster plot of spikes begins to show a zigzag pattern intermingled with inclined partial stripes of spikes due
to local clustering, as shown in Fig.~\ref{fig:NSTDP1}(b2) for $D=150$, and hence the amplitudes of the IPSR $R(t)$ are reduced so much in comparison with those for $D=50$
[see Fig.~\ref{fig:NSTDP1}(c2)]. With further increase in $D$, zigzag stripes in the raster plot are smeared (see the cases of $D=250,$ 350, and 450), and hence the amplitudes of $R(t)$ decrease.
Eventually, when passing $D^*$ desynchronization occurs due to overlap of smeared zigzag stripes, as shown in Fig.~\ref{fig:NSTDP1}(b6) for $D=750$, and then the IPSR $R(t)$ becomes nearly
stationary (i.e., no population rhythm appears) [see Fig.~\ref{fig:NSTDP1}(c6)].

In the synchronized region for $D < D^*$, we also compare population oscillating behaviors of $R(t)$ with spiking behaviors of individual interneurons.
Figure \ref{fig:NSTDP1}(d) shows plots of the population frequency $\langle f_p \rangle_r$ (represented by open circles) and the population-averaged MFR
$\langle \langle f_i \rangle \rangle_r$ (denoted by crosses) of individual interneurons versus $D$.
In each realization, we get $f_p$ from the one-sided power spectrum of $\Delta R(t) (= R(t)- \overline{R(t)})$ (the overbar represents the time average) with mean-squared amplitude normalization
which is obtained from $2^{16}$ data points, and also obtain the MFR $f_i$ for each interneuron through averaging for $2 \times 10^4$ msec; $\langle \cdots \rangle$ denotes a population average over all interneurons.
As examples, power spectra are shown for various values of $D$ in Figs.~\ref{fig:NSTDP1}(e1)-\ref{fig:NSTDP1}(e5).
Moreover, interspike interval (ISI) histograms for individual interneurons are also given in Figs.~\ref{fig:NSTDP1}(f1)-\ref{fig:NSTDP1}(f6).
For each $D$, we obtain the ISI histogram via collecting $10^5$ ISIs from all interneurons; the bin size for the histogram is 0.5 msec.
For $D=50$, $R(t)$ shows a regular oscillation with population frequency $\langle f_p \rangle_r \simeq 63.8$ Hz [see Fig.~\ref{fig:NSTDP1}(e1)].
The ISI histogram in Fig.~\ref{fig:NSTDP1}(f1) has a single peak, and $\langle {\rm ISI} \rangle_r$ (average ISI) $\simeq 15.7$ msec.
Hence, individual interneurons fire regularly like clocks with the population-averaged MFR $\langle \langle f_i \rangle \rangle_r$ which is the same as $\langle f_p \rangle_r$.
For this case, all interneurons make firings in each spiking stripe in the raster plot (i.e., each stripe is fully occupied by spikes of all interneurons).
Consequently, full synchronization with $\langle f_p \rangle_r = \langle \langle f_i \rangle \rangle_r$ occurs for $D=50$. This kind of full synchronization persists until $D=D_{th} \simeq 65$, as shown in Fig.~\ref{fig:NSTDP1}(d). However, when passing the threshold $D_{th}$, full synchronization is developed into sparse synchronization with
$\langle f_p \rangle_r > \langle \langle f_i \rangle \rangle_r$ via a pitchfork bifurcation [see Fig.~\ref{fig:NSTDP1}(d)].

For the case of sparse synchronization, individual interneurons fire at lower rates than the population frequency, and hence only a smaller fraction of interneurons fire in each spiking stripe in the raster plot (i.e., each stripe is sparsely occupied by spikes of a smaller fraction of interneurons). As $D$ is increased, the occupation degree of spikes (representing the density of spiking stripes in the raster plot) decreases, along with decrease in the pacing degree of spikes (denoting smearing of spiking stripes), which results in decrease in amplitudes of $R(t)$ [see Figs.~\ref{fig:NSTDP1}(c2)-\ref{fig:NSTDP1}(c5)].
We note that, with increasing $D$ from $D_{th}$, the interval between stripes in the raster plot becomes smaller. Hence, the population frequency $\langle f_p \rangle_r$ increases monotonically, as shown in Fig.~\ref{fig:NSTDP1}(d) [see also Figs.~\ref{fig:NSTDP1}(e2)-\ref{fig:NSTDP1}(e5)]. As a result, fast sparsely synchronized rhythms appear in the range of $D_{th} < D < D^*$. On the other hand, the population-averaged MFR $\langle \langle f_i \rangle \rangle_r$ is much less than $\langle f_p \rangle_r$ due to sparse occupation of spikes in stripes [see Fig.~\ref{fig:NSTDP1}(d)]. As a result, spiking behaviors of individual interneurons differs markedly from the fast population oscillatory behaviors.

Hereafter, we pay our attention to FSS (occurring for $D_{th} < D < D^*$). Unlike the case of full synchronization, individual interneurons exhibit intermittent spikings phase-locked to the IPSR $R(t)$ at random multiples of the global period $T_G$ of $R(t)$. This ``stochastic phase locking,'' leading to ``stochastic spike skipping,'' is well shown in the ISI histogram with multiple peaks appearing at multiples of $T_G$, as shown in
Figs.~\ref{fig:NSTDP1}(f2)-\ref{fig:NSTDP1}(f5), in contrast to the case of full synchronization with a single-peaked ISI histogram. Similar skipping phenomena of spikings (characterized with multi-peaked ISI histograms) have also been found in networks of coupled inhibitory neurons in the presence of noise where noise-induced hopping from one cluster to another one occurs \cite{GR}, in single noisy neuron models exhibiting stochastic resonance due to a weak periodic external force \cite{Longtin1,Longtin2}, and in inhibitory networks of coupled subthreshold neurons showing stochastic spiking coherence \cite{Kim1,Kim2,Kim3}. Due to this stochastic spike skipping, sparse occupation occurs in spiking stripes in the raster plot. For this case, the ensemble-averaged MFR $\langle \langle f_i \rangle \rangle_r$ of individual interneurons becomes less than the population frequency $\langle f_p \rangle_r$, which results in occurrence of sparse synchronization. We also note that, with increasing $D$, multiple peaks in the ISI histogram begin to merge [see Figs.~\ref{fig:NSTDP1}(f2)-\ref{fig:NSTDP1}(f5)], and hence spiking stripes in the raster plot become more and more smeared (i.e., pacing degree of spikes in the raster plot decreases). Eventually, in the case of desynchronized states for $D>D^*$, multiple peaks overlap completely [e.g., see Fig.~\ref{fig:NSTDP1}(f6) for $D=750$], and hence spikes in the raster plot are completely scattered.

We now measure the degree of FSS in the range of $D_{th} < D < D^*$ by employing the statistical-mechanical spiking measure $M_s$ \cite{RM}.
For the case of FSS, sparse stripes appear successively in the raster plot of spikes. The spiking measure $M_i$ of the $i$th stripe is defined by the product of the occupation degree $O_i$ of spikes
(representing the density of the $i$th stripe) and the pacing degree $P_i$ of spikes (denoting the smearing of the $i$th stripe):
\begin{equation}
M_i = O_i \cdot P_i.
\label{eq:SMi}
\end{equation}
The occupation degree $O_i$ of spikes in the stripe is given by the fraction of spiking neurons:
\begin{equation}
   O_i = \frac {N_i^{(s)}} {N},
\end{equation}
where $N_i^{(s)}$ is the number of spiking neurons in the $i$th stripe.
For the case of sparse synchronization, $O_i<1$, in contrast to the case of full synchronization with $O_i=1$.
The pacing degree $P_i$ of spikes in the $i$th stripe can be determined in a statistical-mechanical way by taking into account their contributions to the macroscopic IPSR $R(t)$.
Central maxima of $R(t)$ between neighboring left and right minima of $R(t)$ coincide with centers of stripes in the raster plot. A global cycle begins from a left minimum of
$R(t)$, passes a maximum, and ends at a right minimum. An instantaneous global phase $\Phi(t)$ of $R(t)$ was introduced via linear interpolation in the region forming a global cycle
(for details, refer to Eqs.~(16) and (17) in \cite{RM}). Then, the contribution of the $k$th microscopic spike in the $i$th stripe occurring at the time $t_k^{(s)}$ to $R(t)$ is
given by $\cos \Phi_k$, where $\Phi_k$ is the global phase at the $k$th spiking time [i.e., $\Phi_k \equiv \Phi(t_k^{(s)})$]. A microscopic spike makes the most constructive (in-phase)
contribution to $R(t)$ when the corresponding global phase $\Phi_k$ is $2 \pi n$ ($n=0,1,2, \dots$). On the other hand, it makes the most destructive (anti-phase) contribution to $R(t)$ when $\Phi_k$
is $2 \pi (n-1/2)$. By averaging the contributions of all microscopic spikes in the $i$th stripe to $R(t)$, we get the pacing degree of spikes in the $i$th stripe (see Eq.~(18) in \cite{RM}):
\begin{equation}
 P_i ={ \frac {1} {S_i}} \sum_{k=1}^{S_i} \cos \Phi_k,
\label{eq:PACING}
\end{equation}
where $S_i$ is the total number of microscopic spikes in the $i$th stripe.
Then, through averaging $M_i$ of Eq.~(\ref{eq:SMi}) over a sufficiently large number $N_s$ of stripes, we get the realistic statistical-mechanical spiking measure $M_s$, based on the IPSR $R(t)$
(see Eq.~(19) in \cite{RM}):
\begin{equation}
M_s =  {\frac {1} {N_s}} \sum_{i=1}^{N_s} M_i.
\label{eq:SM}
\end{equation}
In each realization, we obtain $M_s$ by following $3 \times 10^3$ stripes.

\begin{figure}
\includegraphics[width=0.6\columnwidth]{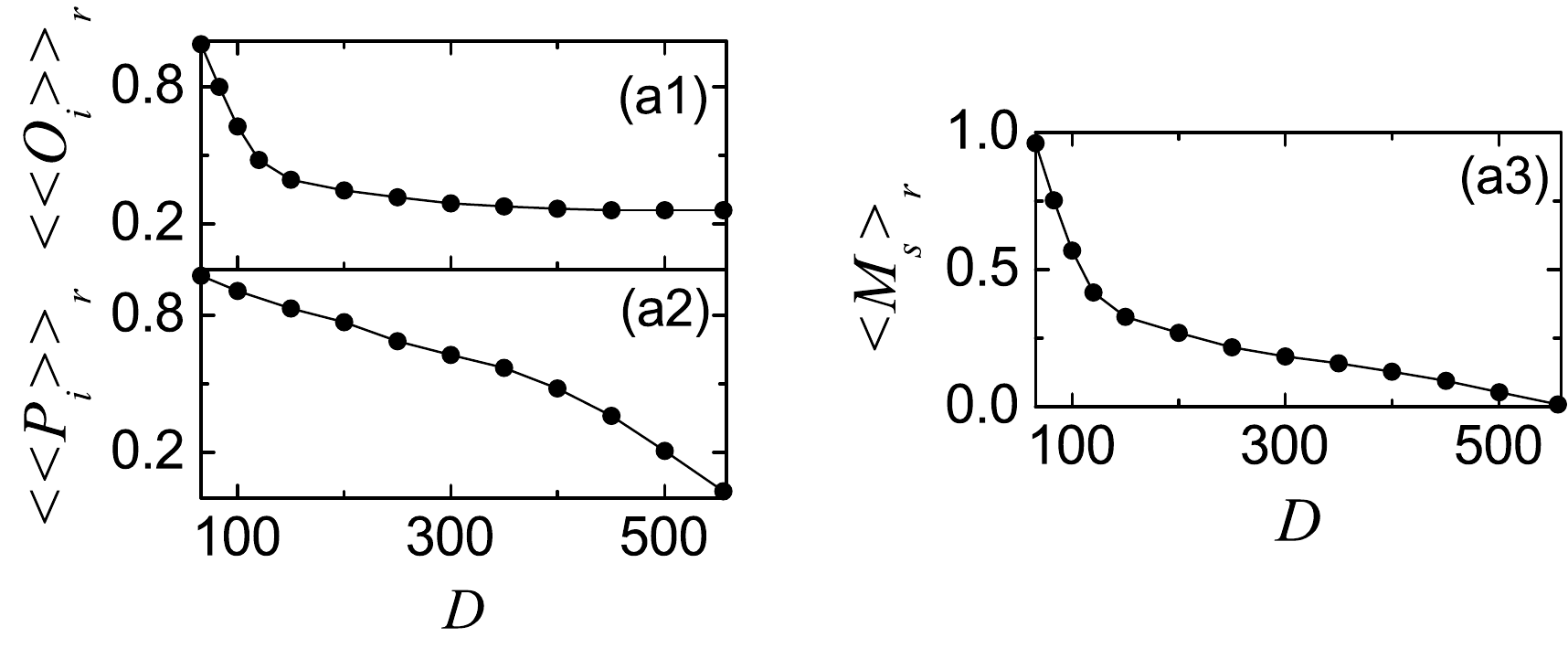}
\caption{Characterization of FSS in the absence of iSTDP.
Plots of (a1) the average occupation degree $\langle \langle O_i \rangle \rangle_r$, (a2) the average pacing degree $\langle \langle P_i \rangle \rangle_r$, and (a3) the statistical-mechanical spiking
measure $\langle M_s \rangle_r$ versus $D$.
}
\label{fig:NSTDP2}
\end{figure}

Figures \ref{fig:NSTDP2}(a1)-\ref{fig:NSTDP2}(a3) show the average occupation degree $\langle \langle O_i \rangle \rangle_r$, the average pacing degree $\langle \langle P_i \rangle \rangle_r$, and
the statistical-mechanical spiking measure $\langle M_s \rangle_r$, respectively. With increasing $D$ from $D_{th}$, at first $\langle \langle O_i \rangle \rangle_r$ (denoting the density of stripes
in the raster plot) decreases rapidly due to stochastic spike skipping, like the behavior of $\langle \langle f_i \rangle \rangle_r$ in Fig.~\ref{fig:NSTDP1}(d), and then it tends to approach a limit value
($\simeq 0.26$). The average pacing degree $\langle \langle P_i \rangle \rangle_r$ represents well the smearing degree of stripes in the raster plot [shown in Figs.~\ref{fig:NSTDP1}(b2)-\ref{fig:NSTDP1}(b5)].
With increasing $D$, $\langle \langle P_i \rangle \rangle_r$ decreases, and for large $D$ near $D^*$ it converges to zero rapidly due to complete overlap of sparse spiking stripes.
Through product of the occupation and the pacing degrees of spikes, the statistical-mechanical spiking measure $\langle M_s \rangle_r$ is obtained. Due to the rapid decrease in
$\langle \langle O_i \rangle \rangle_r$, at first $\langle M_s \rangle_r$ also decreases rapidly, and then it makes a slow convergence to zero for $D=D^*$.

So far, we considered the case of $p=0.25$. From now on, we fix the value of $D$ at $D=350$, and investigate the effect of small-world connectivity on FSS by varying the rewiring probability $p$.
The topological properties of the small-world connectivity has been well characterized in terms of the clustering coefficient and the average path length \cite{SWN1}.
The clustering coefficient $C$, representing the cliquishness of a typical neighborhood in the network, characterizes the local efficiency of information transfer. On the other hand, the average path length $L$, denoting the typical separation between two nodes in the network, characterizes the global efficiency of information transfer.
The Watts-Strogatz SWN interpolates between a regular lattice (corresponding to the case of $p=0$) and a random graph (corresponding to the case of $p=1$) through random uniform rewiring with the probability $p$ \cite{SWN1}.
The regular lattice for $p=0$ is highly clustered but large world where $L$ grows linearly with $N$; $ C \simeq 0.71$ and $L \simeq 17.5$ for $N=10^3$ \cite{SWN1}.
On the other hand, the random graph for $p=1$ is poorly clustered but small world where $L$ grows logarithmically with $N$;
$ C \simeq 0.02$ and $L \simeq 2.64 $ for $N=10^3$ \cite{SWN1}. As soon as $p$ increases from zero, $L$ decreases dramatically, which results in occurrence of a small-world phenomenon which is popularized by the phrase of the ``six degrees of separation'' \cite{SDS1,SDS2}. However, during this dramatic drop in $L$, $C$ decreases only a little. Consequently, for small $p$ SWNs with short path length and high clustering appear (e.g., for $p=0.25$, $C \simeq 0.33$ and $L \simeq 2.83$) \cite{SWN1}.

\begin{figure}
\includegraphics[width=0.8 \columnwidth]{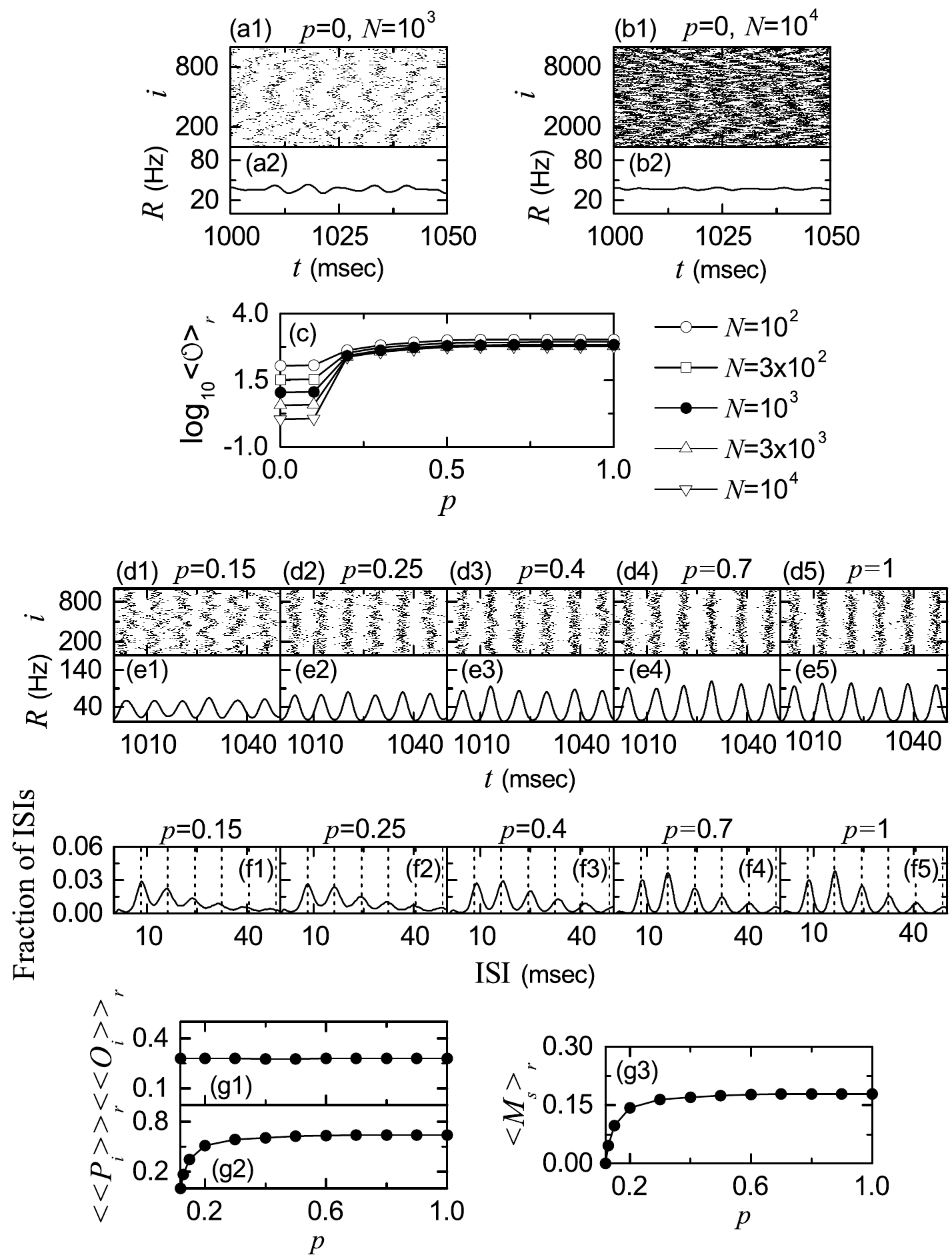}
\caption{
Effect of network architecture on FSS for $D=350$ in the absence of iSTDP.
Desynchronized state for $p=0$: raster plots of spikes in (a1) and (b1) and IPSR kernel estimates $R(t)$ in (a2) and (b2) for $N=10^3$ in (a1) and (a2) and $N=10^4$ in (b1) and (b2).
(c) Plot of the thermodynamic order parameter $\langle {\cal {O}} \rangle_r$ versus $p$. Synchronized states: raster plots of spikes in (d1)-(d5), IPSR kernel estimates $R(t)$ in (e1)-(e5),
and ISI histograms in (f1)-(f5) for various values of $p$. Plots of (g1) the average occupation degree $\langle \langle O_i \rangle \rangle_r$, (g2) the average pacing degree $\langle \langle P_i \rangle \rangle_r$, and (g3)
the statistical-mechanical spiking measure $\langle M_s \rangle_r$ versus $p$.
}
\label{fig:NSTDP3}
\end{figure}

We first consider the population state in the regular lattice for $p=0$. As shown in Fig.~\ref{fig:NSTDP3}(a1) for $N=10^3$, the raster plot shows a zigzag pattern intermingled with inclined partial stripes of spikes, and $R(t)$ is composed of coherent parts with regular large-amplitude oscillations and incoherent parts with irregular small-amplitude fluctuations [see Fig.~\ref{fig:NSTDP3}(a2)]. For $p=0$, the clustering coefficient is high, and hence inclined partial stripes (indicating local clustering of spikes) seem to appear in the raster plot of spikes. As $N$ is increased to $10^4$, partial stripes become more inclined from the vertical [see Fig.~\ref{fig:NSTDP3}(b1)], and hence spikes become more difficult to keep pace with each other. Consequently, $R(t)$ shows noisy fluctuations with smaller amplitudes, as shown in Fig.~\ref{fig:NSTDP3}(b2). Hence, the population state for $p=0$ seems to be desynchronized because $R(t)$ tends to be nearly stationary as $N$ increases to the infinity. With increasing $p$ from 0, long-range short-cuts begin to appear, and hence the average path length $L$ becomes shorter. Hence, when global communication between distant neurons are sufficiently efficient, fast sparsely synchronized states may emerge.
To examine appearance of FSS, we obtain the order parameter $\cal{O}$ of Eq.~(\ref{eq:Order}) by varying $p$. Figure \ref{fig:NSTDP3}(c) shows plots of the order parameter
$\langle {\cal{O}} \rangle_r$ versus $p$. For $p < p^*$ $(\simeq 0.12$), desynchronized states exist because $\langle {\cal{O}} \rangle_r$ tends to zero as $N$ is increased. However, when passing the critical value $p^*$, a transition to FSS occurs because the values of $\langle {\cal{O}} \rangle_r$ become saturated to non-zero limit values.

We now study population and individual behaviors of FSS for $p > p^*$. FSS can be understood well via comparison of population behaviors with individual behaviors.
Figures \ref{fig:NSTDP3}(d1)-\ref{fig:NSTDP3}(d5) and Figures \ref{fig:NSTDP3}(e1)-\ref{fig:NSTDP3}(e5) show raster plots of spikes and IPSR kernel estimates $R(t)$ for various values of $p$, respectively.
As $p$ is increased, the zigzagness degree of partial stripes in the raster plots becomes reduced due to decrease in the clustering coefficient [see Figures~\ref{fig:NSTDP3}(d1)-\ref{fig:NSTDP3}(d5)].
For $p = \tilde{p}$ ($\sim 0.7$), the raster plot becomes composed of vertical stripes without zigzag, and then the pacing degree between spikes for $p > \tilde{p}$ becomes nearly the same.
As a result, the amplitudes of $R(t)$ increase until $p= \tilde{p}$, and then they become nearly saturated. For all these values of $p$, $R(t)$ exhibit regular oscillations with the same population
frequency $f_p \simeq 123$ Hz, corresponding to an ultrafast rhythm ($100-200$ Hz). In contrast to this ultrafast rhythm, individual interneurons show intermittent and stochastic discharges like Geiger counters.
We collect $10^5$ ISIs from all interneurons, and obtain the ISI histograms which are shown in Figs.~\ref{fig:NSTDP3}(f1)-\ref{fig:NSTDP3}(f5).
Individual interneurons show stochastic phase lockings [i.e., intermittent spikings phase-locked to $R(t)$ at random multiples of the global period $T_G$ of $R(t)$], leading to stochastic spike skipping.
Hence, multiple peaks appear at multiples of $T_G$ ($\simeq 8.1$ msec) in the ISI histograms. With increasing $p$, merged multiple peaks begin to be separated, and for $p > \tilde{p}$ nearly the same histograms with clearly separated multiple peaks emerge. Hence, with increasing $p$ the pacing degree between spikes becomes higher, and for $p > \tilde{p}$ it becomes nearly the same.
For all these values of $p$, the population-averaged MFR $\langle f_i \rangle$, corresponding to the inverse of the average ISI, is 34 Hz.
Hence, each interneuron exhibits an average firing sparsely once during about 3.6 population cycles. As a result, for $p>p^*$ fast sparsely synchronized rhythms appear.

For $p>p^*$, we characterize FSS by varying $p$ in terms of the average occupation degree $\langle \langle O_i \rangle \rangle_r$, the average pacing degree $\langle \langle P_i \rangle \rangle_r$, and
the statistical-mechanical spiking measure $\langle M_s \rangle_r$. Figures \ref{fig:NSTDP3}(g1)-\ref{fig:NSTDP3}(g3) show plots of $\langle \langle O_i \rangle \rangle_r$, $\langle \langle P_i \rangle \rangle_r$, and
$\langle M_s \rangle_r$, respectively. The average occupation degree $\langle \langle O_i \rangle \rangle_r$ is nearly the same ($\langle \langle O_i \rangle \rangle_r$ $\simeq 0.28$), independently of $p$; only a fraction
(about 1/3.6) of total interneurons fire in each stripe. This sparse occupation results from stochastic spike
skipping of individual interneurons which is well seen in the multi-peaked ISI histograms. Hence, $\langle \langle O_i \rangle \rangle_r$ characterizes the sparseness degree of FSS well.
In contrast, with increase in $p$, at first the average pacing degree $\langle \langle P_i \rangle \rangle_r$ makes a rapid increase due to appearance of long-range connections.
However, the value of $\langle \langle P_i \rangle \rangle_r$ becomes saturated at $p = \tilde{p}$ because the number of long-range connections which appear up to $\tilde {p}$ is enough to obtain
maximal pacing degree. As in the case of $\langle \langle P_i \rangle \rangle_r$, the statistical-mechanical spiking measure $\langle M_s \rangle_r$ increases rapidly up to $p = \tilde{p}$
because $\langle \langle O_i \rangle \rangle_r$ is nearly independent of $p$. $\langle M_s \rangle_r$ is nearly equal to $\langle \langle P_i \rangle \rangle_r$/3.6 because of
nearly constant sparse occupation [$\langle \langle O_i \rangle \rangle_r$ $\simeq 1/3.6$].

\subsection{Effect of iSTDP on FSS}
\label{subsec:iSTDP}
In this subsection, we study the effect of iSTDP on FSS [occurring for $D_{th} (\simeq 65) < D < D^* (\simeq 558)$ in the absence of iSTDP].
The initial values of synaptic strengths $\{ J_{ij} \}$ are chosen from the Gaussian distribution with the mean $J_0$ (= 700) and the standard
deviation $\sigma_0$ (=5). Then, $J_{ij}$ for each synapse is updated according to a nearest-spike pair-based STDP rule of Eq.~(\ref{eq:MSTDP}).

\begin{figure}[b]
\includegraphics[width=0.9\columnwidth]{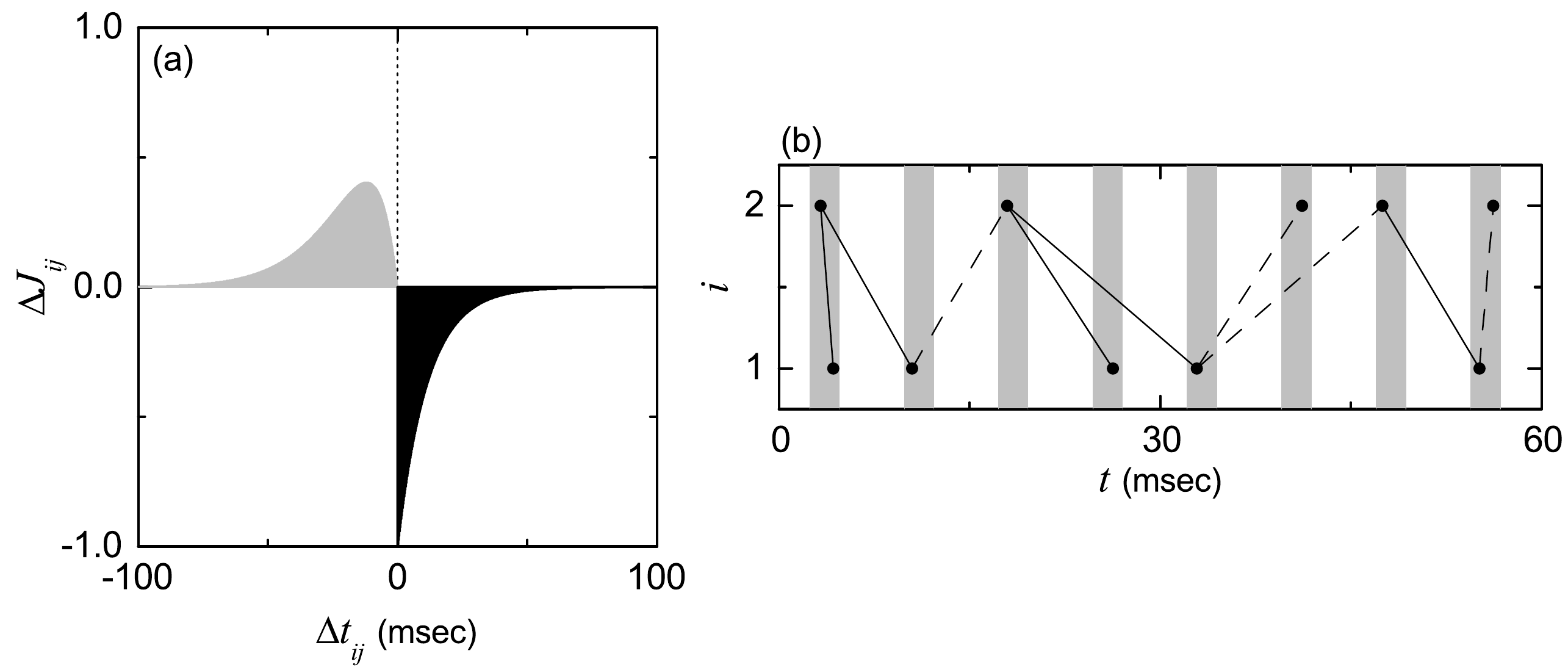}
\caption{(a) Time window for the Anti-Hebbian iSTDP. Plot of synaptic modification $\Delta J_{ij}$ versus $\Delta t_{ij}$ $(=t_i^{(post)} - t_j^{(pre)})$ for $A_+=1$, $A_{-}=1.1$, $\tau_+=11.5$ msec and $\tau_{-}=12$ msec. $t_i^{(post)}$ and $t_j^{(pre)}$ are spiking times of the $i$th post-synaptic and the $j$th pre-synaptic neurons, respectively. (b) Schematic diagram for the nearest-spike pair-based STDP rule; $i=1$ and 2 correspond to the post- and the pre-synaptic neurons. Gray boxes and solid circles denote stripes and spikes, respectively. Solid and dashed lines denote LTD and LTP, respectively.
}
\label{fig:TW}
\end{figure}

Figure \ref{fig:TW}(a) shows an asymmetric anti-Hebbian time window for the synaptic modification $\Delta J_{ij}(\Delta t_{ij})$ of Eq.~(\ref{eq:TW}) versus $\Delta t_{ij}$). $\Delta J_{ij}(\Delta t_{ij})$ varies depending on the relative time difference $\Delta t_{ij}$ $(=t_i^{(post)} - t_j^{(pre)})$ between the nearest spike times of the post-synaptic neuron $i$ and the pre-synaptic neuron $j$.
In contrast to the case of a Hebbian time window for the eSTDP \cite{SSS}, when a post-synaptic spike follows a pre-synaptic spike (i.e., $\Delta t_{ij}$ is positive), LTD of synaptic strength appears; otherwise (i.e., $\Delta t_{ij}$ is negative), LTP occurs. A schematic diagram for the nearest-spike pair-based STDP rule is given in Fig.~\ref{fig:TW}(b), where $i=1$ and 2 correspond to the post- and the pre-synaptic interneurons. Here, gray boxes denote stripes in the raster plot, and spikes in the stripes are denoted by solid circles. When the post-synaptic neuron ($i=1$) fires a spike, LTD (represented by solid lines) occurs through iSTDP between the
post-synaptic spike and the previous nearest pre-synaptic spike. On the other hand, when the pre-synaptic neuron ($i=2$) fires a spike, LTP (denoted by dashed lines) occurs via iSTDP
between the pre-synaptic spike and the previous nearest post-synaptic spike. For the case of sparse synchronization, individual interneurons make stochastic spike skipping (i.e., they make intermittent and stochastic discharges).
As a result of stochastic spike skipping, nearest-neighboring pre- and post-synaptic spikes may appear in any two separate stripes (e.g., nearest-neighboring, next-nearest-neighboring or farther-separated stripes), as well as in the same stripe, in contrast to the case of full synchronization where they appear in the same or just in the nearest-neighboring stripes [compare Fig.~\ref{fig:TW}(b) with Fig. 4(b) (corresponding to the case of full synchronization) in \cite{SSS}]. For simplicity, only the cases, corresponding to the same, the nearest-neighboring, and the next-nearest-neighboring stripes, are shown in Fig.~\ref{fig:TW}(b).

\begin{figure}[b]
\includegraphics[width=0.9\columnwidth]{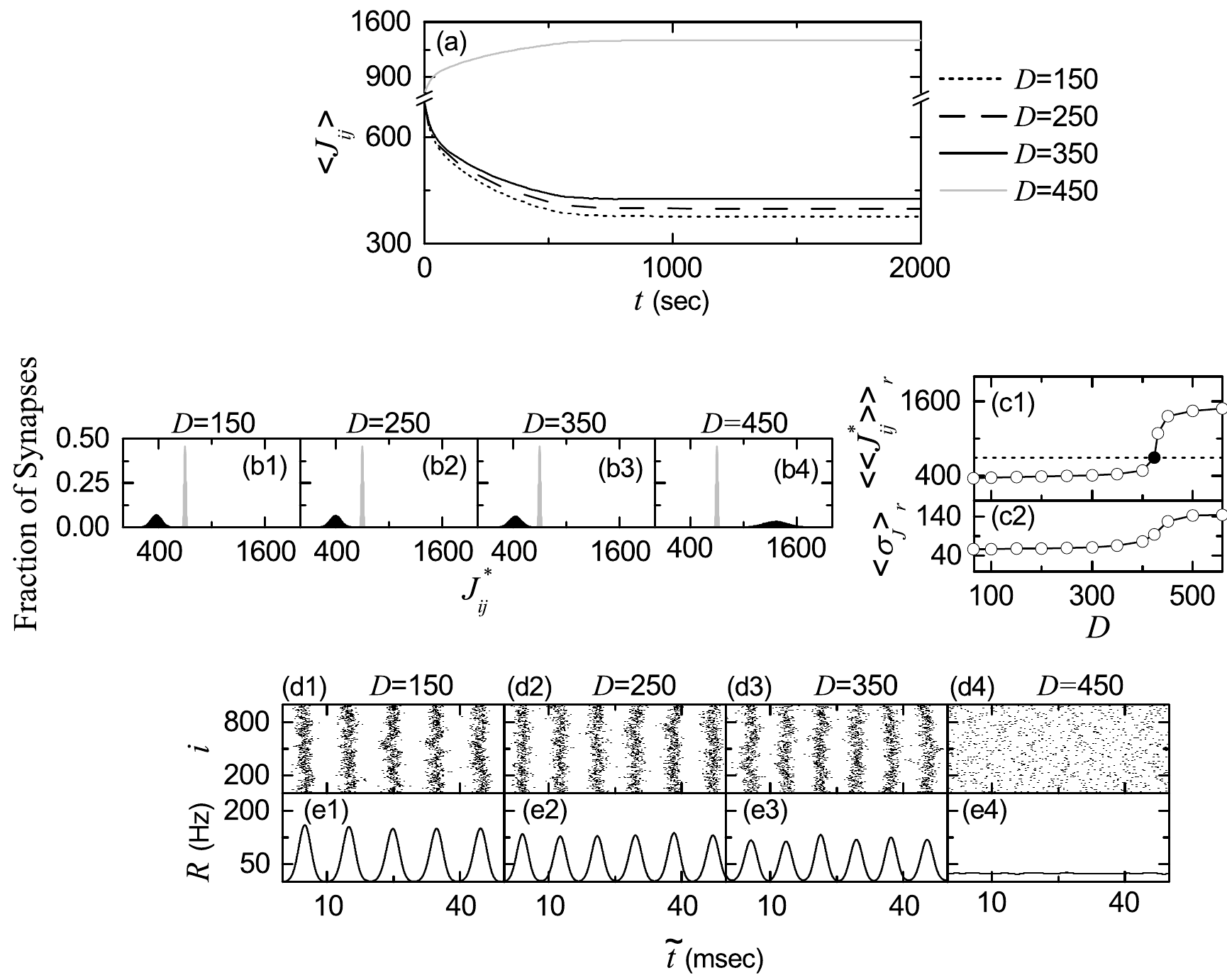}
\caption{Effects of iSTDP on FSS. (a) Time-evolutions of population-averaged synaptic strengths $\langle J_{ij} \rangle$ for various values of $D$.
(b1)-(b4) Histograms for the fraction of synapses versus $J^*_{ij}$ (saturated limit values of $J_{ij}$ at $t = 1000$ sec) are shown in black color for various values of $D$;
for comparison, initial distributions of synaptic strengths $\{ J_{ij} \}$ are also shown in gray color.
Plots of (c1) population-averaged limit values of synaptic strengths $\langle \langle J^*_{ij} \rangle \rangle_r$  (open circles) and (c2) standard deviations $\langle \sigma_J \rangle_r$  (open circles)
versus $D$. Raster plots of spikes in (d1)-(d4) and IPSR kernel estimates $R(t)$ in (e1)-(e4) for various values of $D$ after the saturation time, where
$t=t^*$ (saturation time = 1000 sec) + $\widetilde{t}$.
}
\label{fig:STDP1}
\end{figure}

Figure \ref{fig:STDP1}(a) shows time-evolutions of population-averaged synaptic strengths $\langle J_{ij} \rangle$ for various values of $D$ in the SWN with $p=0.25$; $\langle \cdots \rangle$ represents an
average over all synapses. For each case of $D=150,$ 250, and 350, $\langle J_{ij} \rangle$ decreases monotonically below its initial value $J_0$ (=700), and it approaches a saturated limit value $\langle
J_{ij}^* \rangle$ nearly at $t=1000$ sec. Consequently, LTD occurs for these values of $D$. On the other hand, for $D=450$ $\langle J_{ij} \rangle$ increases monotonically above $J_0$, and
approaches a saturated limit value $\langle J_{ij}^* \rangle$. As a result, LTP occurs for the case of $D=450$.
Histograms for fraction of synapses versus $J_{ij}^*$ (saturated limit values of $J_{ij}$ at $t=1000$ sec) are shown in black color for various values of $D$ in Figs.~\ref{fig:STDP1}(b1)-\ref{fig:STDP1}(b4); the bin size for each histogram is 10. For comparison, initial distributions of synaptic strengths $\{ J_{ij} \}$ (i.e., Gaussian distributions whose mean $J_0$ and standard deviation $\sigma_0$ are 700 and 5, respectively) are also shown in gray color. For the cases of LTD ($D=150,$ 250, and 350), their black histograms lie on the left side of the initial gray histograms, and hence their population-averaged values $\langle J_{ij}^*
\rangle$ become smaller than the initial value $J_0$. On the other hand, the black histogram for the case of LTP ($D=450$) is shifted to the right side of the initial gray
histogram, and hence its population-averaged value $\langle J_{ij}^* \rangle$ becomes larger than $J_0$. For both cases of LTD and LTP, their black histograms are much wider than the initial
gray histograms [i.e., the standard deviations $\sigma_J$ are very larger than the initial one $\sigma_0$]. Figure \ref{fig:STDP1}(c1) shows a plot of population-averaged limit values of synaptic strengths $\langle \langle J_{ij}^* \rangle \rangle_r$ versus $D$. Here, the horizontal dotted line represents the initial average value of coupling strengths $J_0$, and the threshold value $\tilde{D}$ $(\simeq 423)$ for LTD/LTP (where $\langle \langle J_{ij}^* \rangle \rangle_r = J_0$) is represented by a solid circle. Hence, LTD occurs in a larger range of FSS ($D_{th} (\simeq 65) < D < \tilde{D}$); FSS in the absence of iSTDP appears in the range of $D_{th}
< D < D^* (\simeq 558)$. As $D$ is decreased from $\tilde{D}$, $\langle \langle J_{ij}^* \rangle \rangle_r$ decreases monotonically. In contrast, LTP takes place in a smaller range of FSS (i.e., $\tilde{D} < D < D^*$), and
with increasing $D$ from $\tilde{D}$  $\langle \langle J_{ij}^* \rangle \rangle_r$ increases monotonically. Figure \ref{fig:STDP1}(c2) also shows plots of standard deviations $\langle \sigma_J \rangle_r$  versus $D$. All the values of $\langle \sigma_J \rangle_r$ are much larger than the initial values $\sigma_0$ (=5). 
The effects of LTD and LTP on FSS after the saturation time ($t=1000$ sec) may be well shown in the raster plot of spikes and the corresponding IPSR kernel estimate $R(t)$. Figures \ref{fig:STDP1}(d1)-\ref{fig:STDP1}(d4) and Figures \ref{fig:STDP1}(e1)-\ref{fig:STDP1}(e4) show raster plots of spikes and the IPSR kernel estimates $R(t)$ for various values of $D$, respectively. When compared with Figs.~\ref{fig:NSTDP1}(b2)-\ref{fig:NSTDP1}(b5) and Figs.~\ref{fig:NSTDP1}(c2)-\ref{fig:NSTDP1}(c5) in the absence of STDP, the degrees of FSS for the case of LTD ($D=150,$ 250, and 350) are increased (i.e., the amplitudes of $R(t)$ are increased) due to decreased mean synaptic inhibition. On the other hand, in the case of LTP ($D=450$) the population state becomes desynchronized (i.e., $R(t)$ becomes nearly stationary) because of increased mean synaptic inhibition. Due to inhibition, the roles of LTD and LTP in inhibitory synaptic plasticity are reversed in comparison with those in excitatory synaptic plasticity where the degree of population synchronization is increased (decreased) via LTP (LTD) \cite{SSS}.

\begin{figure}
\includegraphics[width=0.9\columnwidth]{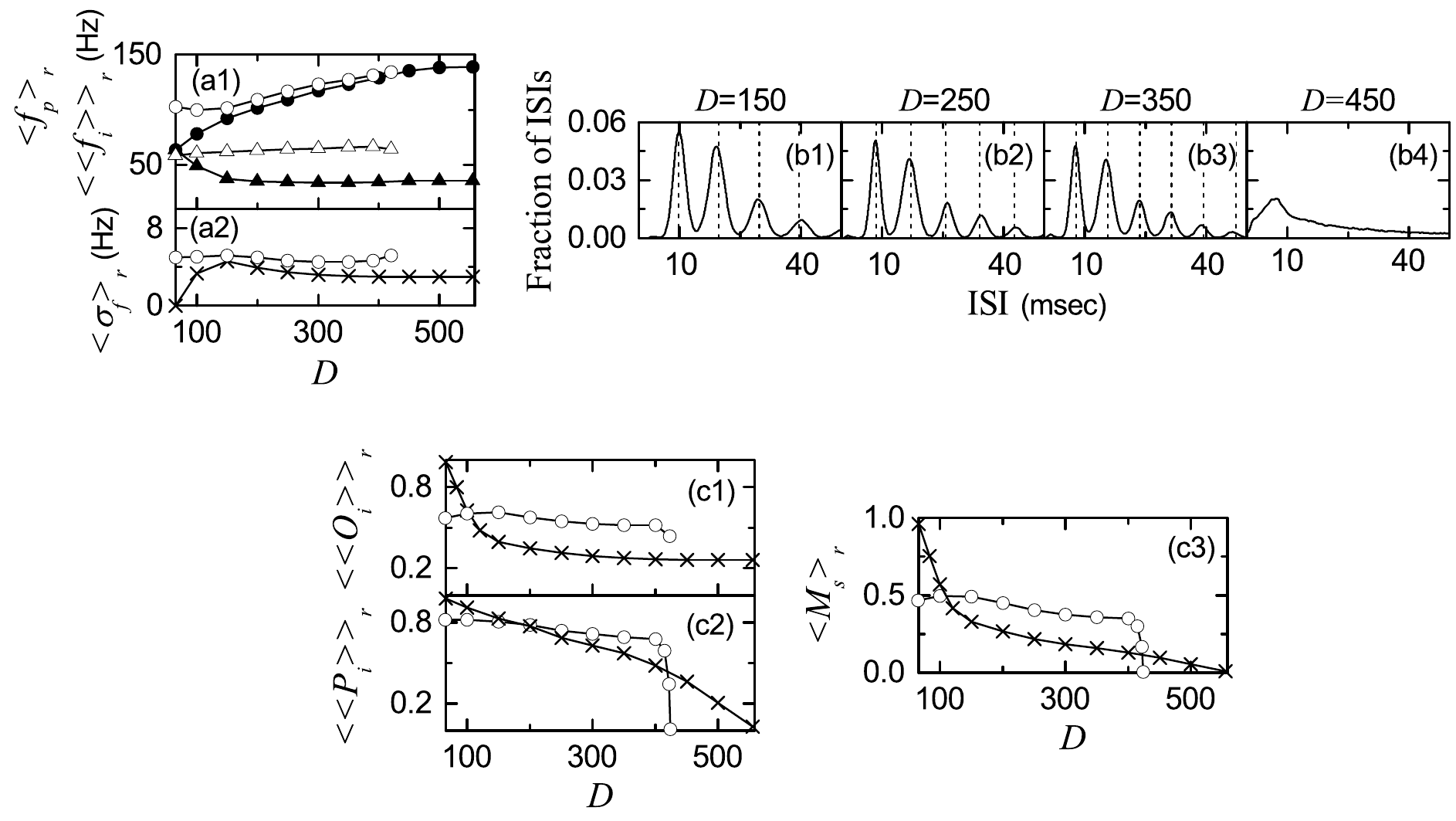}
\caption{Characterization of individual and population behaviors for FSS after the saturation time ($t=1000$ sec) in the presence of iSTDP.
(a1) Plots of population frequency $\langle f_p \rangle_r$ (open circles) and population-averaged MFR $\langle \langle f_i \rangle \rangle_r$ (open triangles); for comparison,
$\langle f_p \rangle_r$ (solid circles) and $\langle \langle f_i \rangle \rangle_r$ (solid triangles) in the absence of iSTDP are also shown.
(a2) Plot of standard deviation $\langle \sigma_{f} \rangle_r$ (open circles) for the distribution of MFRs of individual interneurons versus $D$; crosses represent $\langle \sigma_{f} \rangle_r$ in the absence of iSTDP.
(b1)-(b4) ISI histograms for various values of $D$; vertical dotted lines represent multiples of the global period $T_G$ of the IPSR $R(t)$.
Plots of (c1) the average occupation degree $\langle \langle O_i \rangle \rangle_r$ (open circles), (c2) the average pacing degree $\langle \langle P_i \rangle \rangle_r$ (open circles), and
(c3) the statistical-mechanical spiking measure $\langle M_s \rangle_r$ (open circles) versus $D$. For comparison, $\langle \langle O_i \rangle \rangle_r$, $\langle \langle P_i \rangle \rangle_r$, and $\langle M_s \rangle_r$ in the absence of iSTDP are also denoted by crosses.
}
\label{fig:STDP2}
\end{figure}

In the presence of iSTDP, we also characterize individual and population behaviors for FSS (where individual firing activities differ markedly from population oscillatory behaviors) after the saturation time ($t=1000$ sec)
in the range of $D_{th} (\simeq 65) < D < D^{**} (\simeq 426)$ (where FSS persists in the presence of iSTDP). For comparison, corresponding quantities for FSS in the absence of iSTDP are also
given in the range of  $D_{th} < D < D^{*} (\simeq 558)$ (where FSS appears in the absence of iSTDP).
Figure \ref{fig:STDP2}(a1) shows plots of the population frequency $\langle f_p \rangle_r$ of the IPSR $R(t)$ (open circles) and the population-averaged MFR $\langle \langle f_i \rangle \rangle_r$
of individual interneurons (open triangles) versus $D$; $\langle f_p \rangle_r$ (solid circles) and $\langle \langle f_i \rangle \rangle_r$ (solid triangles) in the absence of iSTDP are also
shown. With decreasing $D$ from $D^*$ to $D_{th}$, $\langle f_p \rangle_r$ and $\langle \langle f_i \rangle \rangle_r$ in the absence of iSTDP approach each other (through decrease in $\langle f_p \rangle_r$
and increase in $\langle \langle f_i \rangle \rangle_r$), and eventually they merge for $D=D_{th}$. As a result, for $D \leq D_{th}$ full synchronization with $\langle f_p \rangle_r = \langle \langle f_i \rangle \rangle_r$
appears in the absence of iSTDP. In the presence of iSTDP, the values of $\langle f_p \rangle_r$ (open circles) are larger than those (solid circles) in the absence of iSTDP mainly due to decreased mean synaptic inhibition $\langle J_{ij} \rangle$ (i.e., LTD). As $D$ is decreased from $D^{**}$, $\langle f_p \rangle_r$ (open circles) decreases. However, from $D \sim 100$ it begins to increase slowly, and then its difference from the value
(solid circles) in the absence of iSTDP increases.
For $D>150$, values of $\langle \langle f_i \rangle \rangle_r$ (open triangles) are much larger than those (solid  triangles) in the absence of iSTDP mainly because of LTD (i.e., decreased mean synaptic inhibition $\langle J_{ij} \rangle$). With decreasing from $D=150$, $\langle \langle f_i \rangle \rangle_r$ (open triangle) begins to decrease slowly, in contrast to increase in $\langle \langle f_i \rangle \rangle_r$ (solid triangle) for the case without iSTDP. For $D \sim 75$ they cross, and then the values of $\langle \langle f_i \rangle \rangle_r$ (open triangles) become a little smaller than those (solid triangles) in the absence of iSTDP. Consequently, for $D=D_{th}$ the difference between $\langle f_p \rangle_r$ (open circle) and $\langle \langle f_i \rangle \rangle_r$ (open triangle) is non-zero (i.e., $\langle f_p \rangle_r > \langle \langle f_i \rangle \rangle_r$), in contrast to the case without iSTDP (where the difference becomes zero). This tendency persists for $D<D_{th}$, and hence full synchronization (with $\langle f_p \rangle_r = \langle \langle f_i \rangle \rangle_r$
for $0 \leq D \leq D_{th}$) in the case without iSTDP breaks up into FSS (with $\langle f_p \rangle_r > \langle \langle f_i \rangle \rangle_r$) in the presence of iSTDP. Figure \ref{fig:STDP2}(a2) also shows a plot of the standard deviation $\langle \sigma_{f} \rangle_r$ (open circles) for the distribution of MFRs of individual interneurons versus $D$; crosses represent $\langle \sigma_{f} \rangle_r$ in the absence of iSTDP. Values of $\langle \sigma_{f} \rangle_r$ (open circles) are larger than those (crosses) in the absence of iSTDP mainly due to increased standard deviation $\langle \sigma_J \rangle_r$ of synaptic strengths. Particularly, near $D_{th}$ their differences become much larger because $\langle \sigma_{f} \rangle_r$ in the absence of iSTDP tends to converge to zero (i.e., $\langle \sigma_{f} \rangle_r=0$ in the case of full synchronization for $D=D_{th}$ without iSTDP). This big difference in $\langle \sigma_f \rangle_r$ near $D_{th}$ in the presence of iSTDP results from the break-up of full synchronization for $D \leq D_{th}$ due to the dominant effect of large dispersions in synaptic inhibition.

For the case of FSS, stochastic phase locking, leading to stochastic spike skipping, is well shown in the ISI histogram with multiple peaks appearing at multiples of the global period $T_G$ of the IPSR $R(t)$, as shown
in Figs.~\ref{fig:STDP2}(b1)-\ref{fig:STDP2}(b3). Due to the stochastic spike skipping, sparse occupation occurs in stripes in the raster plot of spikes. As a result, the ensemble-averaged MFR $\langle f_i \rangle$ of individual interneurons becomes less than the population frequency $f_p$. In comparison with those for the case without iSTDP [see Figs.~\ref{fig:NSTDP1}(f2)-\ref{fig:NSTDP1}(f4)],
the peaks are shifted a little to the left, and the heights of the 1st and the 2nd peaks are increased. Hence, the average ISI $\langle {\rm {ISI}} \rangle$ becomes shorter in the presence of iSTDP, which results
in increase in the population-averaged MFRs $\langle f_i \rangle$. For the cases of $D=350$ and 250, peaks in the presence of iSTDP are clearer than those in the absence of iSTDP, mainly due to decreased
synaptic inhibition $\langle J_{ij} \rangle$ (i.e., LTD), and hence the pacing between spikes in the raster plots are increased for the case with iSTDP. On the other hand, for a smaller case of $D=150$ a little merging between peaks occurs in the presence of iSTDP, mainly because of the dominant effect of increased standard deviation $\sigma_J$ (overcoming the effect of LTD), and hence the pacing between spikes is decreased a little.
For the case of desynchronized states for $D> D^{**}$, complete overlap between multiple peaks occurs [e.g., see the case of $D=450$ in Fig.~\ref{fig:STDP2}(b4)], and hence spikes in the raster plot are completely scattered, as shown in Fig.~\ref{fig:STDP1}(d4).

Figures \ref{fig:STDP2}(c1)-\ref{fig:STDP2}(c2) show the average occupation degree $\langle \langle O_i \rangle \rangle_r$ and the average pacing degree $\langle \langle P_i \rangle \rangle_r$ (represented by open circles), respectively; for comparison, $\langle \langle O_i \rangle \rangle_r$  and $\langle \langle P_i \rangle \rangle_r$ (denoted by crosses) are also shown in the case without iSTDP. In most cases of LTD, the values of
$\langle \langle O_i \rangle \rangle_r$ (open circles) are larger than those (crosses) in the absence of iSTDP, due to decreased mean synaptic inhibition. With decreasing $D$ from $D^{**}$, there are no particular variations in $\langle \langle O_i \rangle \rangle_r$ (open circles). On the other hand, $\langle \langle O_i \rangle \rangle_r$ (crosses) in the absence of iSTDP increases rapidly to 1 near $D_{th}$, because of existence of full synchronization for $D \leq D_{th}$ with $\langle \langle O_i \rangle \rangle_r =1$. Hence, near $D_{th}$, the values (open circles) of $\langle \langle O_i \rangle \rangle_r$ in the presence of iSTDP are smaller than those (crosses) for the case without iSTDP due to the dominant effect of standard deviations of synaptic inhibition strengths (causing break up of full synchronization in the absence of iSTDP).
Next, we consider the average pacing degree $\langle \langle P_i \rangle \rangle_r$. As $D$ is increased from $D_{th}$ to $D^*$, $\langle \langle P_i \rangle \rangle_r$ (crosses) in the absence of iSTDP decreases smoothly.
On the other hand, $\langle \langle P_i \rangle \rangle_r$ in the presence of iSTDP shows a step-like transition. In the region of LTD, there are no particular variations in $\langle \langle P_i \rangle \rangle_r$
(just a little decrease with increasing $D$). Near $\tilde D (\simeq 423)$, a rapid transition to the case of $\langle \langle P_i \rangle \rangle_r =0$ occurs due to LTP (i.e., increased mean synaptic inhibition), in contrast to the smooth decrease in $\langle \langle P_i \rangle \rangle_r$ (crosses) in the absence of iSTDP. For $\sim 200 < D < \tilde{D}$, the values of $\langle \langle P_i \rangle \rangle_r$ (open circles) are larger than those (crosses) in the case without iSTDP mainly because of LTD (i.e., decreased mean synaptic inhibition). However, for $D_{th} < D < \sim 200$ the values (open circles) of $\langle \langle P_i \rangle \rangle_r$ in the presence of iSTDP are smaller than those (crosses) for the case without iSTDP mainly because of the dominant effect of standard deviations of synaptic inhibition strengths.

The statistical-mechanical spiking measure $\langle M_s \rangle_r$ (combining the effect of both the average occupation and pacing degrees) is represented by open circles in Fig.~\ref{fig:STDP2}(c3).
With decreasing from $D^*$ $\langle M_s \rangle_r$ in the absence of iSTDP (denoted by crosses) increases smoothly, and near $D \sim 100$ it begins to increase rapidly due to existence of
full synchronization for $D=D_{th}$. In contrast, in the presence of iSTDP, $\langle M_s \rangle_r$ shows a step-like transition (see open circles). Due to the effect of $\langle \langle P_i \rangle \rangle_r$,
a rapid transition to the case of $\langle M_s \rangle_r =0$ occurs near $\tilde{D}$ because of LTP (decreasing the degree of FSS). On the other hand, in most cases of LTD ($\sim 100 < D < \tilde{D}$),
the values of $\langle M_s \rangle_r$ (open circles) are larger than those (crosses) in the case without iSTDP mainly because of LTD (increasing the degree of FSS). However, for $D_{th} < D < \sim 100$
the values (open circles) of $\langle M_s \rangle_r$ in the presence of iSTDP are smaller than those (crosses) for the case without iSTDP mainly due to the dominant effect of standard deviations of synaptic inhibition strengths (decreasing the degree of FSS). As a result, in most cases of LTD (increasing the degree of FSS), good synchronization with higher $\langle M_s \rangle_r$ gets better; in some other cases near $D_{th}$ the degree of good 
synchronization decreases mainly due to the dominant effect of standard deviation (decreasing the degree of FSS) of synaptic inhibition strengths. On the other hand, in all cases bad synchronization with lower $\langle M_s \rangle_r$ gets worse via LTP (decreasing the degree of FSS). This kind of Matthew effect (valid in most cases of LTD) in inhibitory synaptic plasticity is in contrast to the Matthew effect in excitatory synaptic plasticity where good (bad) synchronization gets better (worse) via LTP (LTD) \cite{SSS,SBS}; the roles of LTD and LTP in the presence of iSTDP are reversed in comparison to those in the case of eSTDP.

\begin{figure}
\includegraphics[width=0.9\columnwidth]{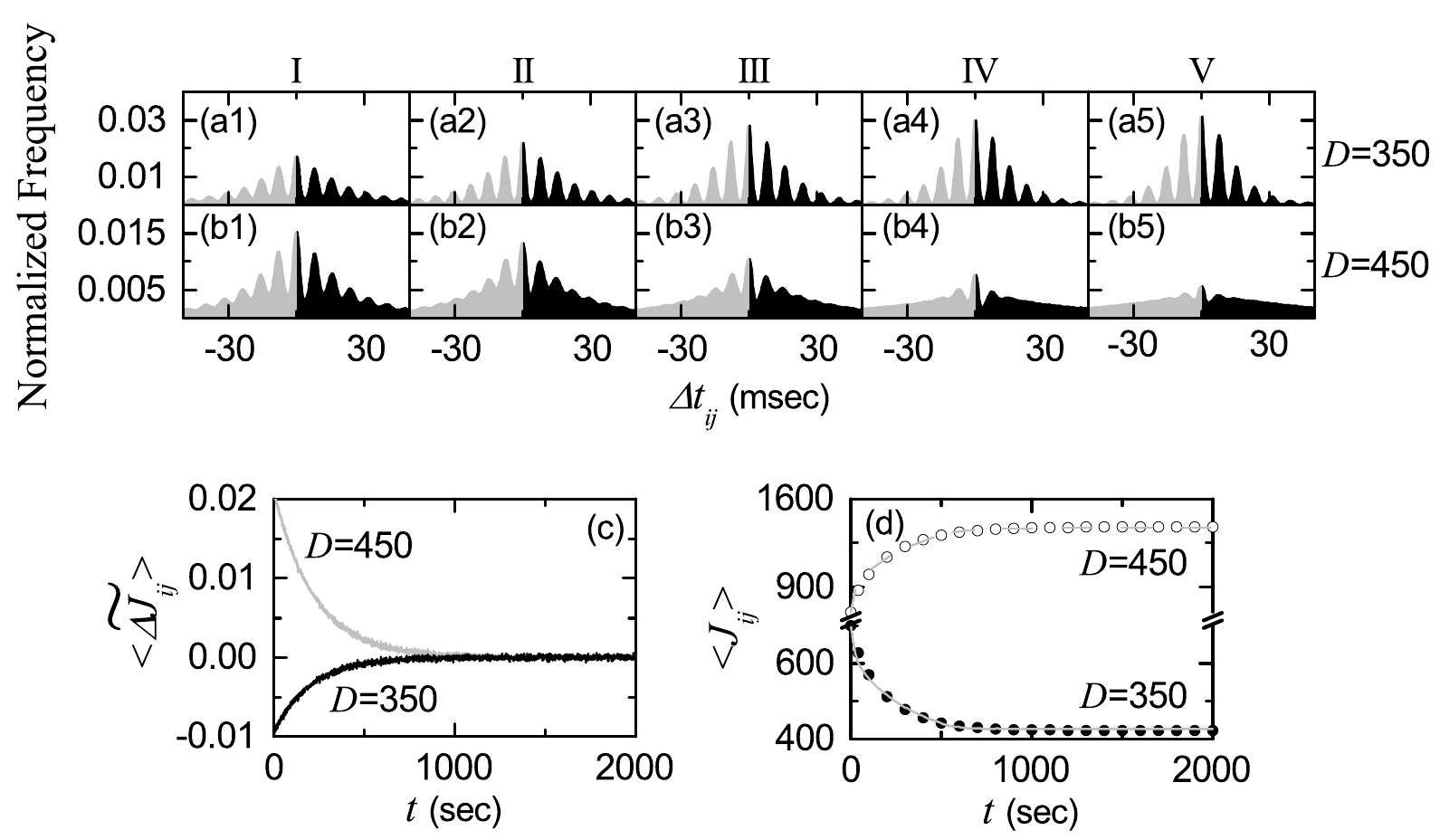}
\caption{Microscopic investigations on emergences of LTD and LTP.
Time-evolutions of the normalized histogram $H(\Delta t_{ij})$ for the distributions of time delays $\{ \Delta t_{ij} \}$ between the pre- and the post-synaptic spike times for $D = 350$ in (a1)-(a5)
and for $D = 450$ in (b1)-(b5); 5 stages are shown in I (starting from 0 sec), II (starting from 100 sec), III (starting from 300 sec), IV (starting from 500 sec), and V (starting from 800 sec).
(c) Time-evolutions of multiplicative synaptic modification $\langle {\widetilde{\Delta J_{ij}}} \rangle$ for $D=350$ (black line) and $D=450$ (gray line).  
(d) Time-evolutions of population-averaged synaptic strength $\langle J_{ij} \rangle$ (obtained by an approximate method) for $D=350$ (solid circle) and $D=450$ (open circle);
gray solid and dashed lines represent ones (obtained by direct calculations) for $D=$ 350 and 450 in Fig.~\ref{fig:STDP1}(a), respectively.
}
\label{fig:STDP3}
\end{figure}

From now on, we make an intensive investigation on emergences of LTD and LTP of synaptic strengths via a microscopic method based on the distributions of time delays $\{ \Delta t_{ij} \}$ between the pre- and
the post-synaptic spike times. Figures \ref{fig:STDP3}(a1)-\ref{fig:STDP3}(a5) and \ref{fig:STDP3}(b1)-\ref{fig:STDP3}(b5) show time-evolutions of normalized histograms $H(\Delta t_{ij})$ for the distributions of time delays $\{ \Delta t_{ij} \}$ for $D=350$ and 450, respectively; the bin size in each histogram is 0.5 msec. Here, we consider 5 stages, represented by I (starting from 0 sec), II (starting from 100 sec), III (starting from 300 sec), IV (starting from 500 sec), and  V (starting from 800 sec). At each stage, we get the distribution of $\{ \Delta t_{ij} \}$ for all synaptic pairs during 0.2 sec and obtain the normalized histogram by dividing the distribution with the total number of synapses (=50000). For $D=350$ (LTD), multi-peaks appear in each histogram, in contrast to the case of full synchronization \cite{SSS}.
As explained in Fig.~\ref{fig:TW}(b), due to stochastic spike skipping, nearest-neighboring pre- and post-synaptic spikes appear in any two separate stripes (e.g., nearest-neighboring, next-nearest-neighboring or farther-separated stripes), as well as in the same stripe, which is similar to the multi-peaked ISI histogram. In the stage I, in addition to the main central (1st-order) peak, higher $k$th-order ($k=2,\dots, 7$) left and right minor peaks also are well seen. Here, LTD and LTP occur in the black ($\Delta t>0$) and the gray ($\Delta t <0$) parts, respectively.
As the time $t$ is increased (i.e., with increase in the level of stage), peaks become narrowed, and then they become sharper. Particularly, heights of major ($k=1,2,3,4$) peaks tend to be increased, while those of minor
($k=5, 6, 7$) peaks seem to be decreased. Intervals between peaks also seem to be decreased a little because the population frequency $f_p$ of $R(t)$ increases a little with the stage.
In the stage I, the effect in the right black part (LTD) is dominant, in comparison with the effect in the left gray part (LTP), and hence the overall net LTD begins to emerge.
As the level of stage is increased, the effect of LTD in the black part tends to nearly cancel out the effect of LTP in the gray part at the stage V.
For $D=450$ (LTP), in the initial stage I, multi-peaks are well seen in the histogram, like the case of $D=350$.
For this initial stage, the effect in the left gray part (LTP) is dominant, in comparison with the effect in the right black part (LTD), and hence the overall net LTP begins to emerge.
However, with increasing the level of stage, peaks become wider and the tendency of merging between the peaks is more and more intensified, in contrast to the case of $D=350$.
Furthermore, the effect of LTP in the gray part tends to nearly cancel out the effect of LTD in the black part at the stage V.

We consider successive time intervals $I_k \equiv (t_{k},t_{k+1})$, where $t_k=0.2 \cdot (k-1)$ sec ($k=1,2,3,\dots$).
With increasing the time $t$, in each $k$th time interval $I_k$, we obtain the $k$th normalized histogram $H_k(\Delta t_{ij})$ ($k=1,2,3,\dots$)
through the distribution of $\{ \Delta t_{ij} \}$ for all synaptic pairs during 0.2 sec.
Then, from Eq.~(\ref{eq:MSTDP}), we get the population-averaged synaptic strength $\langle J_{ij} \rangle_k$ recursively:
\begin{equation}
\langle J_{ij} \rangle_{k} = \langle J_{ij} \rangle_{k-1} + \delta \cdot \langle \widetilde{\Delta J_{ij}}(\Delta t_{ij})  \rangle_{k},
\label{eq:ASS1}
\end{equation}
where $\langle J_{ij} \rangle_0=J_0$ (=700: initial mean value), $\langle \cdots \rangle_k$ means the average over the distribution of time delays $\{ \Delta t_{ij} \}$ for all synaptic pairs in the $k$th
time interval, and the multiplicative synaptic modification $\widetilde{\Delta J_{ij}}(\Delta t_{ij})$ is given
by the product of the multiplicative factor ($J^*-J_{ij}$) [$J_{ij}:$ synaptic coupling strength at the $(k-1)$th stage] and the absolute value of synaptic modification
$| \Delta J_{ij}(\Delta t_{ij}) |$:
\begin{equation}
 \widetilde{\Delta J_{ij}}(\Delta t_{ij})  =  (J^* - J_{ij})~ |\Delta J_{ij}(\Delta t_{ij})|.
\label{eq:ASS2}
\end{equation}
Here, we obtain the population-averaged multiplicative synaptic modification $\langle \widetilde{\Delta J_{ij}}(\Delta t_{ij}) \rangle_{k}$ for the $k$th stage
via a population-average approximation where $J_{ij}$ is replaced by its population average $\langle J_{ij} \rangle_{k-1}$ at the $(k-1)$th stage:
\begin{equation}
 \langle \widetilde{\Delta J_{ij}}(\Delta t_{ij}) \rangle_k  \simeq (J^*- \langle J_{ij} \rangle_{k-1})~ \langle |\Delta J_{ij}(\Delta t_{ij})| \rangle_k.
\label{eq:ASS3}
\end{equation}
Here, $\langle |\Delta J_{ij}(\Delta t_{ij})| \rangle_k$ may be easily obtained from the $k$th normalized histogram $H_k(\Delta t_{ij})$:
\begin{equation}
  \langle |\Delta J_{ij}(\Delta t_{ij})| \rangle_{k}  \simeq  \sum_{\rm bins} H_{k} (\Delta t_{ij}) \cdot | \Delta J_{ij} (\Delta t_{ij}) |.
\label{eq:ASS4}
\end{equation}
Using Eqs.~(\ref{eq:ASS1}), (\ref{eq:ASS3}), and (\ref{eq:ASS4}), we obtain approximate values of $\langle \widetilde{\Delta J_{ij}} \rangle_k$ and $\langle J_{ij} \rangle_{k}$ in a recursive way.
Figure \ref{fig:STDP3}(c) shows time-evolutions of $\langle \widetilde{\Delta J_{ij}} \rangle$ for $D=350$ (black curve) and $D=450$ (gray curve).
$\langle \widetilde{\Delta J_{ij}} \rangle$ for $D=350$ is negative, while $\langle \widetilde{\Delta J_{ij}} \rangle$ for $D=450$ is positive.
For both cases they converge toward nearly zero at the stage V (starting from 800 sec) because the effects of LTD and LTP in the normalized histograms are nearly cancelled out.
The time-evolutions of $\langle J_{ij} \rangle$ for $D=350$ (solid circles) and $D=450$ (open circles) are also shown in Fig.~\ref{fig:STDP3}(d).
We note that the approximately-obtained values for $\langle J_{ij} \rangle$ agree well with directly-obtained ones [denoted by the gray solid (dashed) line for $D=350$ (450)] in Fig.~\ref{fig:STDP1}(a).
Consequently, LTD (LTP) emerges for $D=350$ (450).

\begin{figure}[b]
\includegraphics[width=0.9\columnwidth]{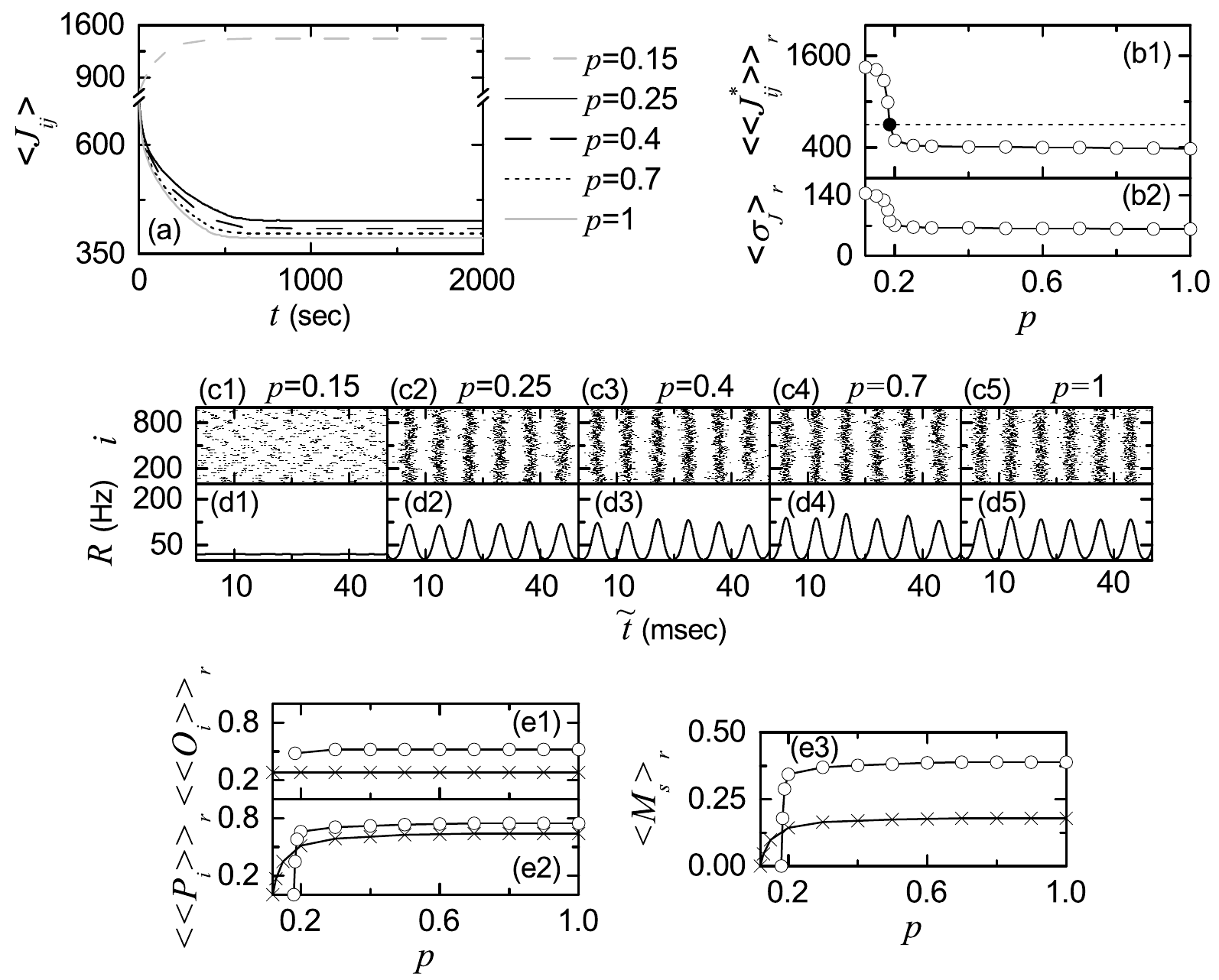}
\caption{Effect of network architecture on FSS for $D=350$ in the presence of iSTDP.
(a) Time-evolutions of population-averaged synaptic strengths $\langle J_{ij} \rangle$ for various values of $p$.
Plots of (b1) population-averaged limit values of synaptic strengths $\langle \langle J^*_{ij} \rangle \rangle_r$  and (b2) standard deviations $\langle \sigma_J \rangle_r$ versus $p$.
Raster plots of spikes in (c1)-(c5) and IPSR kernel estimates $R(t)$ in (d1)-(d5) for various values of $p$.
Plots of (e1) the average occupation degree $\langle \langle O_i \rangle \rangle_r$ (represented by open circles), (e2) the average pacing degree $\langle \langle P_i \rangle \rangle_r$ (denoted by open
circles), and (e3) the statistical-mechanical spiking measure $\langle M_s \rangle_r$ (represented by open circles) versus $p$.
For comparison, $\langle \langle O_i \rangle \rangle_r$, $\langle \langle P_i \rangle \rangle_r$, and $\langle M_s \rangle_r$ in the absence of iSTDP are also denoted by crosses.
}
\label{fig:STDP4}
\end{figure}

Finally, we investigate the effect of network architecture on FSS for $D=350$ by varying the rewiring probability $p$ in the presence of iSTDP;
in the absence of iSTDP, FSS appears for $p > p^* (\simeq 0.12)$. 
Figure \ref{fig:STDP4}(a) shows time-evolutions of population-averaged synaptic strengths $\langle J_{ij} \rangle$ for various values of $p$.
For each case of $p=0.25,$ 0.4, 0.7, and 1.0, $\langle J_{ij} \rangle$ decreases monotonically below its initial value $J_0$ (=700), and it approaches a saturated limit value $\langle
J_{ij}^* \rangle$ nearly at $t=1000$ sec. As a result, LTD occurs for these values of $p$. On the other hand, for $p=0.15$ $\langle J_{ij} \rangle$ increases monotonically above $J_0$, and
approaches a saturated limit value $\langle J_{ij}^* \rangle$. As a result, LTP occurs for the case of $p=0.15$.
Figure \ref{fig:STDP4}(b1) shows a plot of population-averaged limit values of synaptic strengths $\langle \langle J_{ij}^* \rangle \rangle_r$ ($J_{ij}^*$: saturated limit values of $J_{ij}$ at $t=1000$ sec)
versus $p$. Here, the horizontal dotted line represents the initial average value of coupling strengths $J_0$ (= 700), and the threshold value $p_{th}$ $(\simeq 0.185)$ for LTD/LTP (where $\langle \langle J_{ij}^* \rangle \rangle_r = J_0$) is represented by a solid circle. Hence, LTD occurs in a larger range of $p > p_{th}$, while LTP takes place in a smaller range of $p^* < p < p_{th}$.
Figure \ref{fig:STDP4}(b2) also shows a plot of standard deviations $\langle \sigma_J \rangle_r$ versus $p$. All the values of $\langle \sigma_J \rangle_r$ are much larger than the initial value $\sigma_0$ (=5).
The effects of LTD and LTP on FSS after the saturation time ($t=1000$ sec) may be well shown in the raster plot of spikes and the corresponding IPSR kernel estimate $R(t)$ which are shown in
Figs.~\ref{fig:STDP4}(c1)-\ref{fig:STDP4}(c5) and Figs.~\ref{fig:STDP4}(d1)-\ref{fig:STDP4}(d5), respectively. In comparison with Figs.~\ref{fig:NSTDP3}(d1)-\ref{fig:NSTDP3}(d5) and Figs.~\ref{fig:NSTDP3}(e1)-\ref{fig:NSTDP3}(e5) in the absence of STDP, the degrees of FSS for the case of LTD ($p=0.25,$ 0.4, 0.7, and 1.0) are increased (i.e., the amplitudes of $R(t)$ are increased) due to decreased mean synaptic inhibition.
On the other hand, for the case of LTP ($p=0.15$) the population state becomes desynchronized (i.e., $R(t)$ becomes nearly stationary) because of increased mean synaptic inhibition.

Figures \ref{fig:STDP4}(e1) and \ref{fig:STDP4}(e2) show the average occupation degree $\langle \langle O_i \rangle \rangle_r$ and the average pacing degree $\langle \langle P_i \rangle \rangle_r$ of FSS (represented by open circles), respectively; for comparison, $\langle \langle O_i \rangle \rangle_r$  and $\langle \langle P_i \rangle \rangle_r$ (denoted by crosses) are also shown in the case without iSTDP.
In the presence of iSTDP, $\langle \langle O_i \rangle \rangle_r$ (open circles) shows just a little variation, and their values are larger than those (crosses) in the absence of iSTDP, mainly due to LTD.
On the other hand, $\langle \langle P_i \rangle \rangle_r$ in the presence of iSTDP shows a step-like transition.
In most region of LTD, there are no particular variations in $\langle \langle P_i \rangle \rangle_r$ (just a little decrease with decreasing $p$), and their values are larger than those (crosses) in the absence of iSTDP
mainly because of decreased mean synaptic inhibition. However, near $p_{th}$ $(\simeq 0.185)$, a rapid transition to the case of $\langle \langle P_i \rangle \rangle_r =0$ occurs due to LTP (i.e., increased mean synaptic inhibition), in contrast to the smooth decrease in $\langle \langle P_i \rangle \rangle_r$ (crosses) in the absence of iSTDP.
Figure \ref{fig:STDP4}(e3) shows the statistical-mechanical spiking measure $\langle M_s \rangle_r$ (combining the effect of both the average occupation and pacing degrees and represented by open circles) in the range of $p^{**} (\simeq 0.161) < p \leq 1$ (where FSS persists in the presence of iSTDP).
In the absence of iSTDP, with decreasing from $p=1$ to $p^* (\simeq 0.12)$, $\langle M_s \rangle_r$  (denoted by crosses) decreases smoothly. In contrast, in the presence of iSTDP, $\langle M_s \rangle_r$ shows a step-like transition (see open circles). Due to the effect of $\langle \langle P_i \rangle \rangle_r$, a rapid transition to the case of $\langle M_s \rangle_r =0$ occurs near $p_{th}$ because of LTP (decreasing the degree of FSS).
On the other hand, in most region of $p$, the values of $\langle M_s \rangle_r$ (open circles) are larger than those (crosses) in the case without iSTDP, mainly because of LTD (increasing the degree of FSS).
As a result, good synchronization with higher $\langle M_s \rangle_r$ gets better via LTD, while bad synchronization with lower $\langle M_s \rangle_r$ gets worse via LTP. This kind of Matthew effect in inhibitory synaptic plasticity is in contrast to the Matthew effect in excitatory synaptic plasticity where good (bad) synchronization gets better (worse) via LTP (LTD) \cite{SSS,SBS}.

\section{Summary and Discussion}
\label{sec:SUM}
We are interested in synchronized brain rhythms in health and disease \cite{Buz1,TW}. For example, synchronous neural oscillations are used for efficient sensory processing such as binding of the integrated whole image in the visual cortex via synchronization of neural firings \cite{Gray1,Gray2,Singer1,Singer2} In addition to such neural encoding of sensory stimuli, neural synchronization is also correlated with pathological brain rhythms related to neural disease (e.g., Parkinson’s disease, epilepsy, and schizophrenia) \cite{Tass,Grosse,Uhlhaas1,Uhlhaas2}.

Particularly, we are concerned about fast sparsely synchronized rhythms in an inhibitory Watts-Strogatz SWN of Izhikevich FS interneurons. A neural circuit in the major parts of the brain such as thalamus, hippocampus and cortex is composed of a few types of excitatory principal cells and diverse types of inhibitory interneurons. Functional diversity of interneurons increases the computational power of principal cells \cite{Buz2,Buz1}. When the synaptic decay time is enough long, mutual inhibition may synchronize neural firings. By providing a coherent oscillatory output to the principal cells, the interneuronal networks play the role of the backbones (i.e., pacemakers or synchronizers) for many brain rhythms such as the 10-Hz thalamocortical spindle rhythms \cite{GR} and the 40-Hz gamma rhythms in the hippocampus and the cortex \cite{WB,gamma}.
A framework for emergence of sparsely synchronized rhythms (where stochastic and intermittent single-cell firing activity is markedly different from fast population oscillation) was developed in random networks with delayed synaptic connections \cite{Sparse1,Sparse2,Sparse3,Sparse4}. Each interneuron in the interneuronal network receives stochastic external excitatory synaptic inputs. When this background noise is strong, interneurons discharge irregularly as Geiger counters, and the population state becomes desynchronized. However, as the inhibitory recurrent feedback becomes sufficiently strong, the asynchronous state becomes destabilized, and then a synchronized state with irregular and sparse neural discharges appears. In this way, under the balance between strong external random excitation and strong recurrent inhibition, FSS was found to emerge in the interneuronal network.
For this case of FSS, the population frequency $f_p$ is ultrafast (i.e. $100-200$ Hz), while individual interneurons discharge stochastically at much lower rates than $f_p$. This type of fast sparse rhythms was experimentally observed in hippocampal sharp-wave ripples ($100-200$ Hz), associated with memory consolidation, during slow-wave sleep \cite{SWR1,SWR2} and in cerebellar fast oscillations ($\sim 200$ Hz), related to fine motor coordination of inhibitory Purkinje cells \cite{Purkinje1,Purkinje2}.

\begin{figure}
\includegraphics[width=0.9\columnwidth]{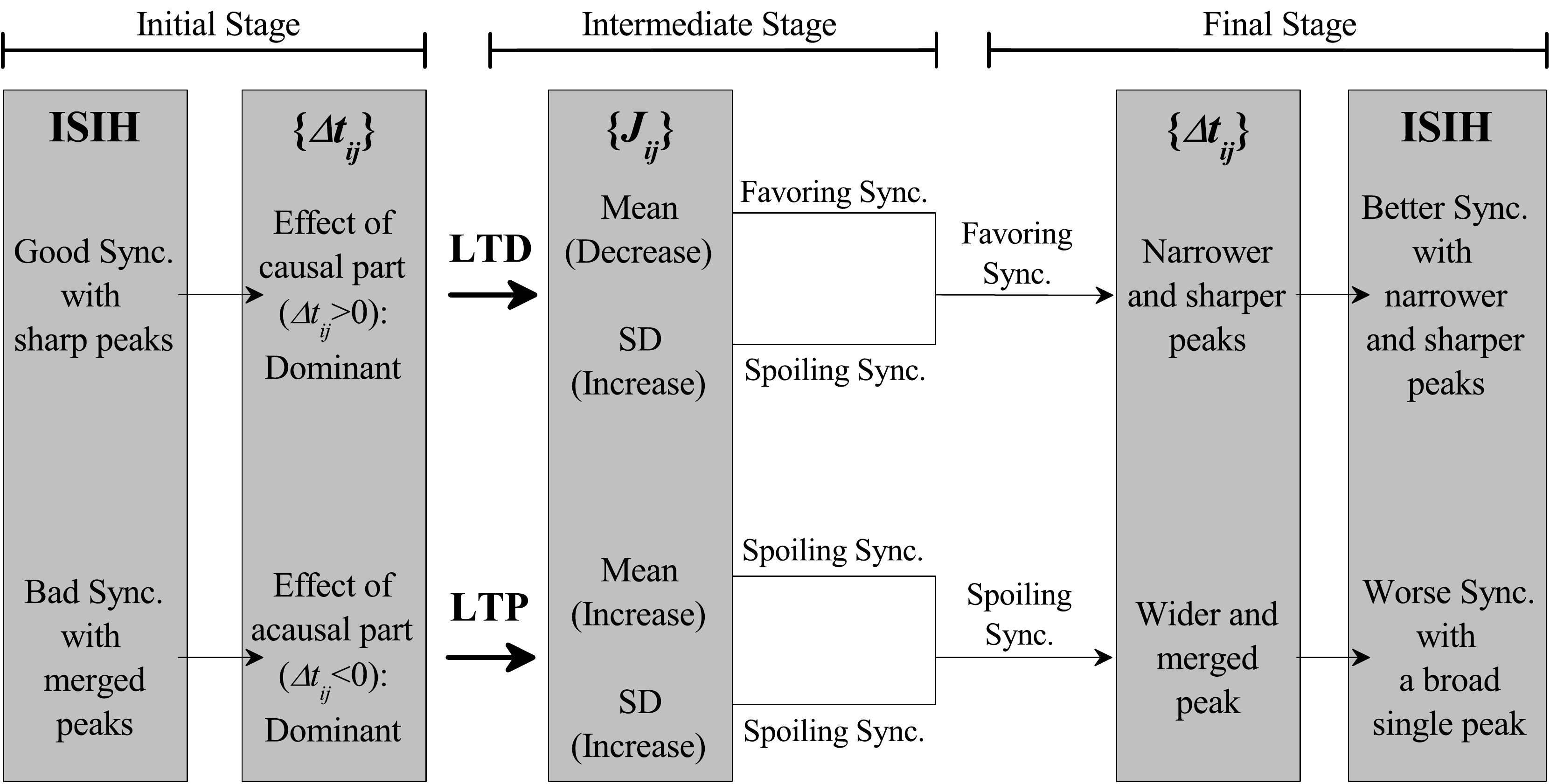}
\caption{Summary on the effects of iSTDP on good and bad synchronization: time-evolutions of good and bad synchronization in the presence of iSTDP. Here, synchronization and standard deviation are abbreviated as sync. and SD, respectively.
}
\label{fig:STDP5}
\end{figure}

In previous works on FSS, synaptic inhibition strengths were static (i.e., inhibitory synaptic plasticity was not considered). On the other hand, in the present work, adaptive dynamics of synaptic inhibition strengths are governed by the iSTDP (which controls the efficacy of diverse computational functions of interneurons). The effects of iSTDP on FSS have been investigated in the SWN with $p=0.25$ by varying the noise intensity $D$. An asymmetric anti-Hebbian time window has been used for the iSTDP update rule, in contrast to the Hebbian time window for the case of eSTDP.
Our results on the effects of iSTDP on FSS are well summarized in the diagram in Fig.~\ref{fig:STDP5}. For the case of FSS, the ISI histogram consists of multiple peaks, due to stochastic spike skipping. These multiple peaks are sharp for the case of good synchronization (with higher spiking measure), while they are merged in the case of bad synchronization (with lower spiking measure) (see the 1st column in Fig.~\ref{fig:STDP5}). Emergences of LTD and LTP of synaptic inhibition strengths were investigated via a microscopic method based on the distributions of time delays $\{ \Delta t_{ij} \}$ between the nearest spiking times of the pre- and the post-synaptic interneurons. Like the case of multi-peaked ISI histogram, sharp multi-peaks appear in the normalized histogram for the distribution of $\{ \Delta t_{ij} \}$; the heights of peaks for the case of good synchronization are higher than those in the case of bad synchronization. For the case of good synchronization, the effect of causal part with $\Delta t>0$ is dominant, and hence LTD begins to occur (see the 2nd column in Fig.~\ref{fig:STDP5}). On the other hand, in the case of bad synchronization the effect of acausal part with $\Delta t <0$ is dominant, and hence LTP begins to take place.

The distribution of synaptic inhibition strengths $\{ J_{ij} \}$ is evolved in the presence of iSTDP (see the 3rd column in Fig.~\ref{fig:STDP5}). For the case of good synchronization, its mean $\langle J_{ij} \rangle$ is decreased (i.e., LTD occurs), while in the case of bad synchronization $\langle J_{ij} \rangle$ is increased (i.e., LTP occurs). The standard deviations $\sigma_J$ for both cases of good and bad synchronization increase. Decrease (increase) in the mean
$\langle J_{ij} \rangle$ [i.e., LTD (LTP)] favors (disfavors) FSS due to increased (decreased) population-averaged MFR $\langle f_i \rangle$ of individual interneurons. Increased standard deviation $\sigma_J$ leads to increase in variation of inhibitory synaptic inputs to individual interneurons, and hence distributions of MFRs of individual interneurons become broader (i.e., standard deviation $\sigma_f$ for the distribution of MFRs increases). Due to  increased $\sigma_f$, it becomes difficult for interneurons to keep their pacing, which results in decrease in the degree of FSS (i.e., FSS is spoiled), as in the case of increasing the noise intensity D. In this way, dispersion of synaptic inhibition strengths seems to play a role which is similar to that of noise.

For the case of iSTDP, in addition to the effect of mean value (LTP or LTD), the effect of standard deviation $\sigma_J$ on population synchronization may also become significant in some cases, in contrast to the case of eSTDP where the mean of LTP/LTD was found to be always dominant \cite{SSS,SBS}. For most cases of good synchronization, the effect of LTD (increasing the degree of FSS) is dominant in comparison with the effect of standard deviation $\sigma_J$ (decreasing the degree of FSS). Consequently, in most cases of good synchronization, it has been found to get better via LTD; in some other cases where the effect of standard deviation $\sigma_J$ is dominant (occurring near $D_{th}$), the degree of good synchronization decreases even in the presence of LTD. In contrast, for all cases, bad synchronization has been found to get worse via LTP (decreasing the degree of FSS). This type of Matthew effect (valid in most cases of LTD) in inhibitory synaptic plasticity is in contrast to the Matthew effect in excitatory synaptic plasticity \cite{SSS,SBS}. For the case of eSTDP, LTP (LTD) has a tendency to favor (disfavor) synchronization via positive (structural) feedback, while LTD (LTP) for the case of iSTDP has a tendency favoring (disfavoring) FSS through a negative (structural) feedback. Hence, due to inhibition via negative feedback, the roles of LTD and LTP in inhibitory plasticity are reversed in comparison with those in excitatory synaptic plasticity through positive feedback where good (bad) synchronization gets better (worse) via LTP (LTD). Consequently, in most region of LTD, the degree of FSS becomes increased, and a rapid transition from FSS to desynchronization occurs via LTP, in contrast to the relatively smooth transition in the absence of iSTDP.

The process of iSTDP may be well visualized in the normalized histogram of $H(\Delta t_{ij})$ and the ISI histogram (see the 4th and 5th columns in Fig.~\ref{fig:STDP5}). With increasing time $t$, peaks in the normalized histogram $H(\Delta t_{ij})$ becomes narrowed and sharper for the case of LTD, while in the case of LTP peaks become wider and merged. After a sufficient time, the effect of LTD in the right causal part with $\Delta t>0$ nearly cancels out the effect of LTP in the left acausal part with $\Delta t <0$. Then, saturated limit states appear without further change in synaptic strengths. For most cases of good synchronization, peaks in the ISI histogram become clearer (i.e., narrower and sharper), mainly due to the dominant effect of decreased synaptic inhibition (i.e., LTD), and hence the pacing between spikes in the raster plot is increased. As a result, in most cases good synchronization gets better via LTD. In contrast, for the case of bad synchronization, complete overlap between peaks in the ISI histogram occurs (i.e., a broad single peak appears) due to increased synaptic inhibition (i.e., LTP), and hence spikes are completely scattered in the raster plot. Consequently, bad synchronization gets worse via LTP.

Emergences of LTD and LTP of synaptic inhibition strengths were investigated via a microscopic method based on the distributions of time delays $\{ \Delta t_{ij} \}$ between the nearest spiking times of the pre- and the post-synaptic interneurons. Time evolutions of normalized histograms $H(\Delta t_{ij})$ were followed for both cases of LTD and LTP. Using a recurrence relation, we recursively obtained population-averaged synaptic inhibition strength $\langle J_{ij} \rangle$ at successive stages through an approximate calculation of population-averaged multiplicative synaptic modification $\langle \widetilde{\Delta J_{ij}} \rangle$ of Eq.~(\ref{eq:ASS3}), based on the normalized histogram at each stage. These approximate values of $\langle J_{ij} \rangle$ have been found to agree well with directly-calculated ones. Consequently, one can understand clearly how microscopic distributions of $\{ \Delta t_{ij} \}$ contribute to $\langle J_{ij} \rangle$ or more directly to $\langle \widetilde{\Delta J_{ij}} \rangle$.

By varying the rewiring probability $p$ in the SWN, we also studied the effect of network architecture on FSS in the presence of iSTDP. As in the above case of variation in $D$ for $p=0.25$, a Matthew effect has also been found to occur in the case of variation in $p$ for $D=350.$ As a result, good (bad) synchronization with higher (lower) spiking measure $M_s$ for $p > (<) p_{th}$ ($\simeq 0.185$) gets better (worse) via LTD (LTP).

Finally, we discuss limitations of our work and future works.
In our work, we employed the standard ``duplet'' STDP model, based on the nearest pre- and post-synaptic spike pairs. However, unfortunately this pair-based STDP model accounts for neither the dependence of plasticity on the repetition frequency of the pairs of pre- and post-synaptic spikes, nor the results of recent triplet and quadruplet experiments \cite{TriSTDP1,TriSTDP2}. Hence, as a future work, it would be interesting to study the effect of iSTDP on FSS by using a triplet iSTDP rule and to compare its results with those for the case of duplet iSTDP rule.
Inhibitory neurons have been found to possess diverse types of plasticity rules. For example, Hebbian STDP \cite{Kullmann,EtoI2}, anti-Hebbian STDP \cite{Kullmann,iSTDP2,EtoI1,EtoI3}, and anti-Hebbian STDP with only LTD \cite{EtoE7} were observed in the case of excitatory (E) to inhibitory (I) connection.
Moreover, for the case of I to E connection, anti-symmetric Hebbian STDP \cite{iSTDP10} and symmetric (non-Hebbian) STDP \cite{iSTDP11} were found. However, in the preset work, for simplicity we assumed that all interneurons exhibit identical anti-Hebbian STDP for the case of I to I connection. To take into consideration heterogeneity on synaptic plasticity of inhibitory cells seems to be beyond the present work, and it will be left as a future work.
In the present work, we considered only the interneuronal network. As explained above, a major neural circuit consists of two excitatory and inhibitory populations.
In previous works \cite{Sparse3,Sparse4}, they also considered the two-population network with four (I to I, I to E, E to I, and E to E) types of connections. The additional I to E, E to I, and E to E connections have tendency to decrease the population frequency $f_p$. It was found that $f_p$ was much reduced to about $30-100$ Hz (corresponding to gamma rhythms), when compared with the case of pure interneuronal network. Hence, in future, it would be interesting to study the effects of interpopulation (I to E and E to I) STDP on FSS in the two-population network of inhibitory Izhikevich FS interneurons and excitatory Izhikevich regular-spiking neurons, in addition to the studied intrapopulation (E to E and I to I) STDP.
In our work, we also considered just the spiking neurons. In addition to spiking, bursting is also another type of neuronal firing activities. Burstings occur when neuronal activity alternates, on a slow timescale, between a silent phase and an active (bursting) phase of fast repetitive spikings. There are several representative examples of bursting neurons.
Recently, we also investigated the effect of iSTDP on burst synchronization in a scale-free neuronal network of inhibitory Hindmarsh-Rose bursting neurons \cite{BSiSTDP}. Thus, the effect of iSTDP on burst synchronization was also found to be in contrast to the effect of eSTDP on burst synchronization, similar to the case of spiking neurons.
Finally, we note that there exist some limitations to STDP in views of biological contexts \cite{Markram}. For example, STDP has been questioned as a general model of synaptic plasticity \cite{Lisman}, the classic STDP windows for LTP and LTD were found to be only one of many possible ones \cite{DynHebb}, and because of the attenuation of the back-propagating action potential, STDP was found to depend on the dendritic synapse location
\cite{Froemke,SynLocation1,SynLocation2}. In the presence of these limitations, we expect that our results on inhibitory synaptic plasticity of I to I connections would make some contributions for understanding the effects of iSTDP on fast sparsely synchronized rhythms

\section*{Acknowledgments}
This research was supported by the Basic Science Research Program through the National Research Foundation of Korea (NRF) funded by the Ministry of Education (Grant No. 20162007688).

\end{document}